\documentclass[fleqn,usenatbib]{mnras}

\usepackage{newtxtext,newtxmath}

\usepackage[T1]{fontenc}

\DeclareRobustCommand{\VAN}[3]{#2}
\let\VANthebibliography\thebibliography
\def\thebibliography{\DeclareRobustCommand{\VAN}[3]{##3}\VANthebibliography}

\usepackage{graphicx}	
\usepackage{amsmath}	
\usepackage{amssymb}	
\usepackage{subfigure}

\usepackage{tikz,xcolor,hyperref}

\definecolor{lime}{HTML}{A6CE39}
\DeclareRobustCommand{\orcidicon}{
	\begin{tikzpicture}
	\draw[lime, fill=lime] (0,0) 
	circle [radius=0.16] 
	node[white] {{\fontfamily{qag}\selectfont \tiny ID}};
	\draw[white, fill=white] (-0.0625,0.095) 
	circle [radius=0.007];
	\end{tikzpicture}
	\hspace{-2mm}
}

\foreach \x in {A, ..., Z}{\expandafter\xdef\csname orcid\x\endcsname{\noexpand\href{https://orcid.org/\csname orcidauthor\x\endcsname}
			{\noexpand\orcidicon}}
}

\title[Schwarzschild Modeling of Barred Galaxy NGC 4371]{Schwarzschild Modeling of Barred S0 Galaxy NGC 4371}

\author[B. Tahmasebzadeh et al.]{
            Behzad Tahmasebzadeh,$^{1,2}$ \thanks{Corr author: behzad@umich.edu;} 
		Ling Zhu,$^{1}$              \thanks{Corr author: lzhu@shao.ac.cn;} 
		Juntai Shen,$^{3,4}$       \thanks{Corr author: jtshen@sjtu.edu.cn.} 
            Dimitri A. Gadotti,$^{5,6}$
            Monica Valluri,$^{2}$
            \newauthor
            Sabine Thater, $^{7}$
            Glenn van de Ven, $^{7}$
            Yunpeng Jin,$^{1}$
            Ortwin Gerhard,$^{8}$
            Peter Erwin,$^{8}$
            Prashin Jethwa,$^{7}$
            \newauthor
            Alice Zocchi,$^{7}$
            Edward J. Lilley,$^{7}$
            Francesca Fragkoudi,$^{9}$
            Adriana de Lorenzo-Cáceres,$^{10,11}$
            \newauthor
            Jairo Méndez-Abreu,$^{10,11}$
            Justus Neumann,$^{12}$
            Rui Guo,$^{3,4}$ 
		\\ 
            \\
            $^{1}$ Shanghai Astronomical Observatory, Chinese Academy of Sciences, 80 Nandan Road, Shanghai 200030, China \\
		$^{2}$ Department of Astronomy, University of Michigan, Ann Arbor, MI, 48109, USA \\
		$^{3}$ Department of Astronomy, School of Physics and Astronomy, Shanghai Jiao Tong University, 800 Dongchuan Road, Shanghai 200240, China \\
		$^{4}$ Key Laboratory for Particle Astrophysics and Cosmology (MOE) / Shanghai Key Laboratory for Particle Physics and Cosmology, Shanghai 200240, China \\
		$^{5}$ European Southern Observatory, Karl-Schwarzschild-Str. 2, D-85748 Garching, Germany \\
		$^{6}$ Centre for Extragalactic Astronomy, Department of Physics, Durham University, South Road, Durham DH1 3LE, UK \\
            $^{7}$ Department of Astrophysics, University of Vienna,
              T\"urkenschanzstraße 17, 1180 Vienna, Austria \\
            $^{8}$ Max-Planck-Institut f{\"u}r Extraterrestrische Physik, Gie{\ss}enbachstra{\ss}e 1, 85748 Garching, Germany \\
            $^{9}$ Institute for Computational Cosmology, Department of Physics, Durham University, DH1 3LE, United Kingdom \\
            $^{10}$ Instituto de Astrofísica de Canarias, C/ Vía Láctea S/N E-38205 La Laguna, Spain \\
            $^{11}$ Departamento de Astrofísica, Universidad de La Laguna, Avda. Astrofísico Francisco Sánchez, E-38205 La Laguna, Spain \\
            $^{12}$ Max-Planck-Institute for Astronomy, Königstuhl 17, D-69117 Heidelberg, Germany \\
	}

\begin{document}
	\label{firstpage}
	\pagerange{\pageref{firstpage}--\pageref{lastpage}}
	\maketitle
   
	\begin{abstract}
We apply the barred Schwarzschild method developed by \citet{behzad.2022} to a barred S0 galaxy, NGC 4371, observed by IFU instruments from the TIMER and ATLAS3D projects. We construct the gravitational potential by combining a fixed black hole mass, a spherical dark matter halo, and stellar mass distribution deprojected from  $3.6$ $\mu$m S$^4$G image considering an axisymmetric disk and a triaxial bar. We independently modelled kinematic data from TIMER and ATLAS3D. Both models fit the data remarkably well. We find a consistent bar pattern speed from the two sets of models with $\Omega_{\rm p} =  23.6 \pm 2.8 \hspace{.08cm} \mathrm{km \hspace{.04cm} s^{-1} \hspace{.04cm} kpc^{-1} }$ and $\Omega_{\rm p} =  22.4 \pm 3.5 \hspace{.08cm} \mathrm{km \hspace{.04cm} s^{-1} \hspace{.04cm} kpc^{-1} }$, respectively. The dimensionless bar rotation parameter is determined to be $ \mathcal{R} \equiv R_{\rm cor}/R_{\rm bar}=1.88 \pm 0.37$, indicating a likely slow bar in NGC 4371.  Additionally, our model predicts a high amount of dark matter within the bar region ($M_{\rm DM}/ M_{\rm total}$ $\sim 0.51 \pm 0.06$), which, aligned with the predictions of cosmological simulations, indicates that fast bars are generally found in baryon-dominated disks. Based on the best-fitting model, we further decompose the galaxy into multiple 3D orbital structures, including a BP/X bar, a classical bulge, a nuclear disk, and a main disk. The BP/X bar is not perfectly included in the input 3D density model, but BP/X-supporting orbits are picked through the fitting to the kinematic data. This is the first time a real barred galaxy has been modelled utilizing the Schwarzschild method including a 3D bar.
	
\end{abstract}


	
\section{Introduction}
\par
About two-thirds of disk galaxies in the local universe host bars \cite[]{Eskridge.2000, Erwin.2018}. These bars are identified either by non-axisymmetric features in their surface density or through kinematic signatures, such as a positive correlation between the mean velocity and the third Gauss-Hermite moment $ h_{3} $ \cite[]{2005.Bureau, 2018.Li}.
\par

Bars can play a significant role in the formation of galaxies by redistributing the energy and the angular momentum of the disk materials \cite[]{1998.Debattista, 2003.Athanassoula, KK.2004, 2011.Gadotti}. Numerous observations and simulations have demonstrated that bars are often associated with a boxy/peanut or X-shaped structure (hereafter BP/X) when viewed edge-on \cite[]{Combes.1981,Raha.1991, Ltticke.2000}. 

\par
The key parameters characterizing a bar include its radius, strength, and pattern speed (e.g., \cite{Aguerri.2015}). The radius and strength of a bar are typically derived from optical or near-infrared images \cite[]{Aguerri.1998, Buta.2001}.  In contrast, the bar pattern speed is a dynamical parameter that is more challenging to measure, requiring kinematic data. \cite{Tremaine.1984} introduced a simple and model-independent method (TW method) for measuring the bar pattern speed $\Omega_{\rm p} $, which is widely used. The TW method uses the profiles of surface brightness $\Sigma (x)$ and line-of-sight (LOS) velocity $V_{los}$ measured along the slits crossing the bar and parallel to the disk major axis, with the coordinate of $x$ integrated from $- \infty$ to $\infty$ along a slit.
\par
In recent decades, integral field unit (IFU) surveys such as SAURON \cite[]{Bacon.2010}, CALIFA \cite[]{CALIFA.2012}, SAMI \cite[]{SAMI.2012}, and MaNGA \cite[]{MANGA.2015}, have provided kinematic maps of thousands of nearby galaxies. 
IFU data improve the accuracy of the pattern speed measurement. The TW method has been applied to sub-samples of barred galaxies from CALIFA \cite[]{Aguerri.2015, Cuomo.2019}, MaNGA \cite[]{Guo.2019, Garma.2022, Tobias.2023}, and MUSE observed galaxies \cite[]{Cuomo.2019b, Buttitta.2022, Cuomo.2022}. The accuracy of the TW method depends on accurately determining the disk position angle. Inaccuracies of a few degrees in the disk position angle can lead to errors of 10\% (and up to 100\%) for $\Omega_{\rm p}$ \cite[]{Debattista.2003, Zou.2019}. There is still large uncertainty in the pattern speed measurements obtained from MaNGA-like data due to their low spatial resolution. MUSE instrument \cite[]{Bacon.2010} offers IFU data with higher spatial resolution and signal-to-noise ratio (S/N). However, accurately measuring $\Omega_{\rm p}$ requires kinematic data covering both the bar and the disk's outer regions. Such observations with MUSE are expensive \cite[]{Cuomo.2019b, Buttitta.2022, Cuomo.2022}. 
\par

Recently, the TIMER project \cite[]{TIMER.2020} observed a sample of 21 nearby barred galaxies using the MUSE instrument. These observations uncovered various structures co-existing with bars in the galaxy centres, such as classical bulges, nuclear disks, and ring-like structures.
To thoroughly understand the formation of these structures, it is crucial to decompose them both morphologically and kinematically to quantify their contributions. Additionally, the bar pattern speed for the TIMER galaxies cannot be determined using the TW method, as the kinematic data are limited only to the bar region.

Dynamical modelling is a powerful method that can constrain the bar pattern speed using full kinematic information and facilitate the dynamical decomposition of the bar, classical bulge, and nuclear disk structures. The bar pattern speed of the Milky Way was strongly constrained by a few dynamical models, including the \cite{Schwarzschild.1979} orbit-superposition method \cite[]{1996.Zhao,  Wang.2012, Wang.2013} and Made-to-Measure method \cite[]{Long2013, Portail.2017a}. The orbit-superposition method, in particular, the \cite{bosch.2008} triaxial code (hereafter \texttt{VdB08}) has been widely used in exploring stellar orbit distribution in a large sample of galaxies from surveys like CALIFA \citep{Zhu2018b, ling.1018}, MaNGA \citep{Jin.2020}, and SAMI \citep{Giulia.2022}. It has been further developed to include the stellar age and metallicity \citep{Zhu.2020,Poci2019}, enabling chemo-dynamical decomposition of galaxy structures \citep{Zhu.2022, Ding.2023}. 
A new version of the \texttt{VdB08} code, named DYNAMITE, has been publicly released, featuring ongoing enhancements \cite[]{Jethwa.2020,Sabine.2022}.


\par
However, the dynamical modelling of external barred galaxies is challenging due to their complicated morphological and kinematic properties. \textit{N}-body simulations are used as input of the 3D density distribution in the previous barred models, like the triaxial bulge/bar/disk M2M model for M31 \cite[]{blana.2018}, and Schwarzschild FORSTAND code \cite[]{Vasiliev.2019c}. Recently, \cite{Dattathri.2023} introduced a new method that employs a parametric 3D density distribution to deproject edge-on barred galaxies with BP/X-shaped structures. This approach was validated using dynamical modelling against mock data with the FORSTAND code.

In \cite{Behzad.2021}, we presented a deprojection method to estimate the 3D density distribution of barred galaxies across various observational orientations, incorporating both an axisymmetric disk and a triaxial (predominantly prolate) bar. Subsequently, we utilized these 3D density distributions as input and modified the \texttt{VdB08} code to explicitly include the bar \citep{behzad.2022}. Testing our methodology with a set of mock data in various orientations has demonstrated its proficiency in accurately recovering key properties of barred galaxies, particularly the bar pattern speed and the BP/X structure.

\par
In this study, we apply our bar modelling approach to NGC 4371, a particularly intriguing barred galaxy observed by the TIMER and ATLAS3D projects. This marks the first application of the Schwarzschild method to a real barred galaxy, with the bar explicitly included in the model. NGC 4371 is notable for its complex inner structures, including a nuclear disk and a bar. There is ongoing debate regarding the presence of a classical bulge in this galaxy \citep{Erwin.2015, Gadotti.2015}. The stellar population across the galaxy appears to be very old, and the bar pattern speed remains undetermined due to limited data coverage. By developing an orbit-superposition model, we aim to first constrain key parameters such as the bar pattern speed, and then quantitatively determine the contributions of the classical bulge, bar, and nuclear disk.
\par
The paper is organized as follows. In Section \ref{S:data}, we introduce the photometry and spectroscopy data used for modelling. In Section \ref{S:method}, we describe the bar modelling steps and the technical details.  In Section \ref{S:results}, we present our results and highlight the key properties of NGC 4371 that have been measured. Section \ref{S:con} provides a summary and conclusion. The Appendix discusses how our new approach improves upon previous axisymmetric models applied to a large sample of spiral galaxies without including a bar.


\section{DATA}\label{S:data}
\subsection{General Properties of NGC 4371}
\par
NGC 4371, a massive early-type galaxy with a stellar mass of $ M_{\star} \sim 10^{10.5} \, M_{\odot}$ \cite[]{Mateos.2015}, is located near the centre of the Virgo cluster at a distance of approximately 16.9 Mpc \cite[]{Blakeslee.2009}. The inner region of this galaxy is composed of a few different structures as seen from photometric images and kinematic maps. It is commonly accepted that NGC 4371 has a bar \citep{Erwin.1999, Buta.2015}. Morphological studies using photometry suggest that it might have a pseudobulge or a composite bulge, including a small classical merger-built bulge alongside a pseudobulge \citep{Fisher.2010, Erwin.2015}. While 2D stellar kinematics do not provide direct evidence for a classical bulge, they instead strongly demonstrate the existence of a relatively large, rapidly rotating nuclear stellar disk extending to $\sim 12^{''}$, which corresponds to the barlens observed in the photometric images \citep{Gadotti.2015, Gadotti.2019}. It is difficult to clearly identify the presence or absence of a BP/X bulge in NGC 4371 from morphological isophotos \citep{Erwin.2013}, although with a high probability of presenting one at its stellar mass \citep{Erwin.2023}. Both the photometric images and stellar kinematic maps are with information blended along the line-of-sight. Considering the complicated composition of a few structures in this galaxy, uncovering its 3D structure combining photometric and kinematic maps might be a key to further understanding its bulge properties. 
\par

\subsection{Photometry}\label{S:photometry} 
We use the $3.6$ $\mu$m image of NGC 4371 taken by the Infrared Array Camera (IRAC) Channel 1 from the Spitzer Survey of Stellar Structures in Galaxies (S$^4$G) \cite[]{Sheth.2010}. The pixel size of the S$^4$G image is $0.75^{''}$ and the point spread function (PSF) FWHM is $\approx 1.8^{''}$ \cite[]{Kim.2014}. Due to the reduced effects of dust extinction and emission at these wavelengths, the S$^4$G image effectively represents the galaxy stellar structures.

 
\subsection{Spectroscopy}\label{S:kin}
\par
MUSE has observed NGC 4371 as part of the TIMER project \cite[]{Gadotti.2019}. MUSE covers an almost square $ 1^{'} \times 1^{'}$ field of view (FOV) with contiguous sampling of $ 0.2^{''} \times 0.2^{''}$ and the spectral coverage of $4750-9350$ $ \rm \mathring{A} $. The spectral sampling is $1.25$ $\rm \AA$ per pixel, and the total integration time is $3840$ s. The stellar kinematics maps of NGC 4371 had been studied in \cite{Gadotti.2015, TIMER.2020}. 
\par
To achieve our desired number of stellar kinematic constraints, we re-extract the 2D maps of $v$, $\sigma$, $h_3$, and $h_4$ from the MUSE data cubes using the Penalized Pixel-Fitting (pPXF) software \cite[]{Cappellari.2004} through the GIST pipeline \cite[]{Bittner.2019},  We spatially binned the spectra to achieve a minimum signal-to-noise (S/N) ratio of approximately 140 per spectral pixel. This was done using the Voronoi binning technique presented by \cite{Cappellari.2003}. We set a minimum signal-to-noise (S/N) threshold of  $\sim 3$ for each pixel adopted for the binning, thereby reducing contamination from outer regions with very low S/N. 

\begin{figure*}
	\centering	%
	\includegraphics[width=1.97\columnwidth]{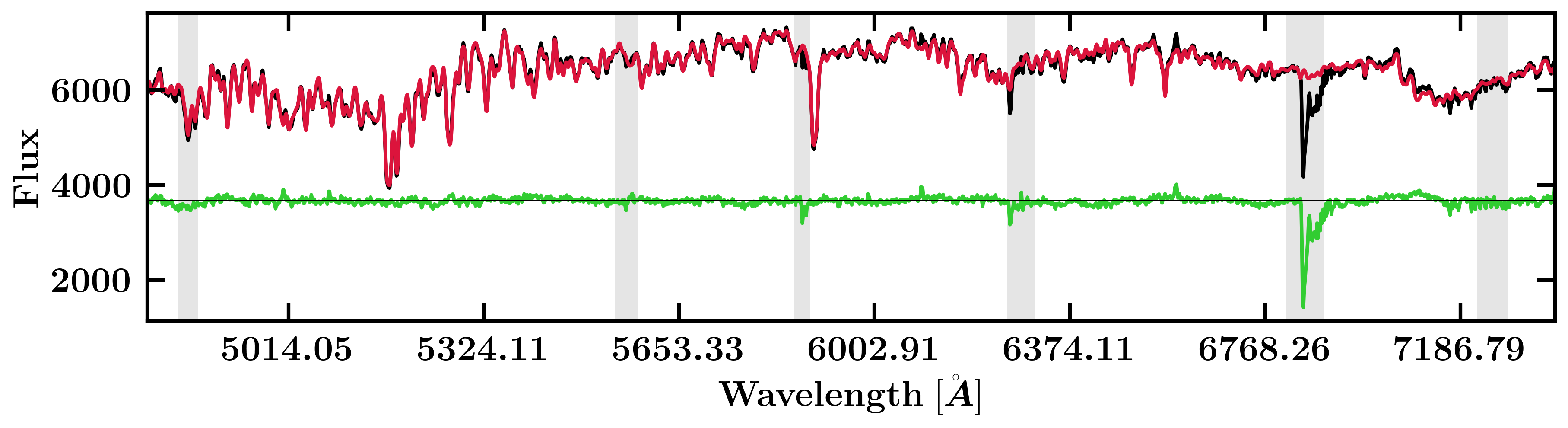}
         \includegraphics[width=2\columnwidth]{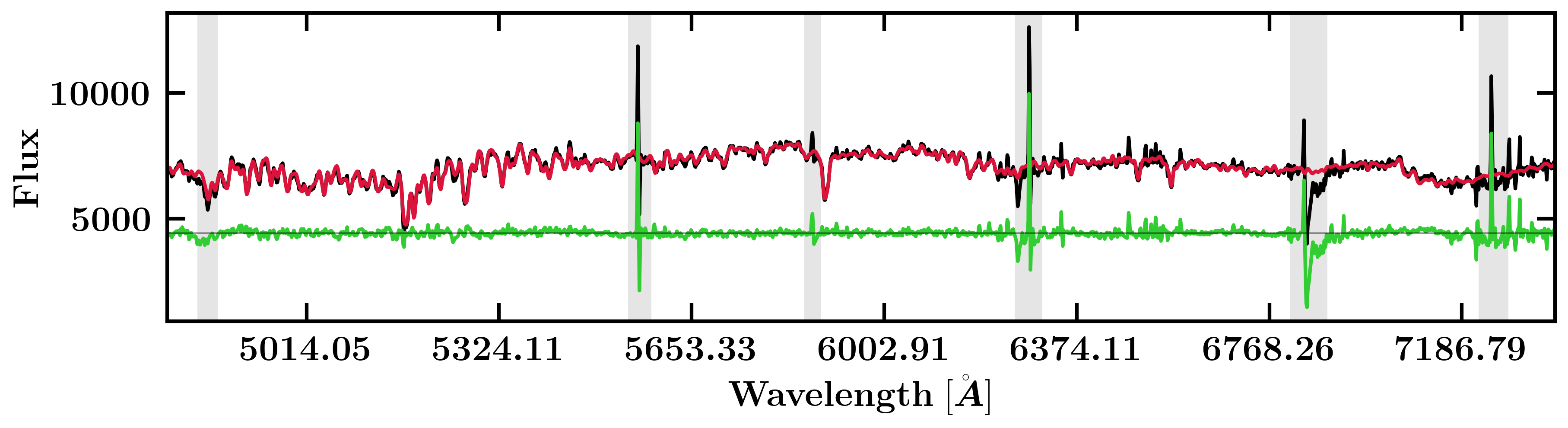}
	\caption{Integrated MUSE spectra (black line) and pPXF fits of NGC 4371 (red lines) are shown for two Voronoi bins, one from the central region (top) and one from the outer region (bottom). The spectra are plotted over the wavelength range of $4800-7300$ $ \rm \mathring{A} $, fitted to the MILES stellar templates. The residuals between the observed spectrum and the best-fitting model are displayed as a green line. Grey-shaded areas indicate regions that were masked during the fit, often due to emission lines or insufficient sky subtraction.} %
	\label{fig:ppxf}%
\end{figure*}

The stellar templates are taken from the Medium-resolution Isaac Newton Telescope Library of Empirical Spectra (MILES) stellar library \cite[]{blzquez.2006, Barroso.2011}. We used the full sample consisting of 980 stars that span the wavelength range of $4760-7400$ $ \rm \mathring{A} $. For our analysis, we specifically fitted the galaxy spectrum within the range of $4800-7300$ $ \rm \mathring{A} $ to ensure compatibility with the coverage provided by the MILES spectral library. Additionally, in the outer region bins, the noise level significantly increases beyond 7300 $ \rm \mathring{A} $ due to weaker signals combined with the higher sky background. We also employed a non-constant Line Spread Function (LSF) to account for the wavelength dependence of the instrumental spectral resolution. We adopted the LSF derived by \cite{Bacon.2017}, where the FWHM varies from 2.98 $ \rm \mathring{A} $ at a wavelength of 4800 $ \rm \mathring{A} $ to 2.54 $ \rm \mathring{A} $ at a wavelength of 7300 $ \rm \mathring{A} $. We used the instrumental dispersion of the MILES template library, reported as 2.51 $ \rm \mathring{A} $  by \cite{Beifiori.2011, Barroso.2011}. \cite{Thater.2019} demonstrated that the effect of the spectral resolution variation, changing from 2.5 to 2.9 $ \rm \mathring{A} $ on the extracted velocity dispersion is on average only 5 km/s, which is within the range of kinematic errors. A $\chi^{2}$ minimization was used with pPXF to fit the stellar template to the spectra from each Voronoi bin. We adopted an 8th-order multiplicative polynomial for the fitting process and a 4th-order additive Legendre polynomial to account for the underlying continuum. Emission lines and regions with poor sky subtraction were masked during the fit. We then compared the fitted spectrum with the original spectrum for each bin. The standard deviation of the residuals is shown as a green line in Fig. \ref{fig:ppxf} for two Voronoi bins from the central and outer regions. The extracted kinematic maps are overall consistent with the results from \cite{TIMER.2020}.
 \par
NGC 4371 is also observed with SAURON \cite[]{Bacon.2001} as part of the volume-limited ATLAS3D project that examined stellar and gas kinematics and photometric imaging of 260 early-type galaxies \cite[]{Cappellari.2011}. The SAURON $ 33^{''} \times 41^{''}$ FOV was sampled by $ 0.94^{''} \times 0.94^{''}$ square lenslets in the low resolution mode. Therefore, SAURON covers a smaller inner region of NGC 4371 than the MUSE datacube. We use the stellar kinematic datacube provided on the ATLAS3D website \footnote{\url{http://www-astro.physics.ox.ac.uk/atlas3d/}} which is extracted with the spectral coverage of $4800-5400$ $ \rm \mathring{A} $.

\par

\section{DYNAMICAL MODEL}\label{S:method}

\subsection{Gravitational Potential}\label{S:pot}

We assume the gravitational potential includes contributions from stellar mass, dark matter (DM), and a fixed central black hole (BH) mass. As the TIMER-MUSE kinematic data do not resolve the BH sphere of influence, we cannot constrain the BH mass in our modelling. Therefore, we fixed the BH mass with $M_{\rm BH} = 10^{6.8} M_{\odot}$ for NGC 4371 constrained by the SINFONI data with higher spatial resolution \citep{Saglia.2016}. Including or excluding the BH with this mass does not affect our modelling outcome. However, its presence provides additional stability in orbit integration in the very central region and helps achieve a more realistic representation of the orbital structure in the very inner region. Despite this, its overall effect on the dynamical properties of the modelled galaxy is negligible. The central black hole can influence stellar orbits beyond its sphere of influence, such as reducing the population of bar-supporting resonances with the smallest pericenter by up to $\sim 15\%$ depending on the BH mass. However, this effect is still confined to a very small region, possibly a few times the sphere of influence, and will not impact the overall projected kinematics in the centre as discussed in \cite{Vance.2023}.




\subsubsection{Dark matter Mass}

For the DM distribution, we consider a spherical Navarro–Frenk–White (NFW) \cite[]{NFW.1996} halo for simplicity and to reduce the number of free parameters using the concentration-mass relation. 
Our kinematic data have limited spatial coverage, and what the model constrained is only the enclosed mass profile within the data coverage; the halo profile beyond the data coverage does not affect the
fitting to the kinematic data. With the current data, an NFW halo with a fixed mass-concentration relation has enough freedom to represent the enclosed DM profile within the data coverage. 
We have tested different dark-matter profiles against the mock model presented in \cite{behzad.2022} and found no significant differences in the modelling results. This is a widely adopted and accepted assumption in stellar dynamical modelling of external galaxies with IFU data covering the inner regions (e.g., \cite{Cappellari.2013, Zhu2018b, Jin.2020, Giulia.2022}, among others).

The enclosed mass profile in the NFW halo can be expressed as: 
\begin{equation}\label{DM:NFW}
M(<r)=M_{200} g(c)\left[\ln \left(1+c r / r_{200}\right)-\frac{c r / r_{200}}{1+c r / r_{200}}\right],
\end{equation}
where $g(c)=[\ln (1+c)-c /(1+c)]^{-1}$ and $ c $ represents the concentration of the DM halo. The virial mass $M_{200}$, is defined as $\frac{4}{3} \pi 200 \rho_{\mathrm{c}} r_{200}^{3}$ representing the mass within the virial radius $r_{200}$. The adopted critical density is $\rho_{\mathrm{c}}=1.37 \times 10^{-7} M_{\odot} \mathrm{pc}^{-3}$, so that the two remaining free parameters are the concentration $c$ and the virial mass $ M_{200} $.
\par
Since the data do not extend to a sufficiently large radius, we cannot constrain $c$ and $M_{200}$ at the same time. Therefore, we fix $c$ based on the relation from \citep[]{Dutton.2014}:%
\begin{equation}\label{DM:c}
\log _{10} c=0.905-0.101 \log _{10}\left(M_{200} /\left[10^{12} h^{-1} M_{\odot}\right]\right),
\end{equation}
which is inferred from galaxy simulations with $h = 0.671$ \citep[]{plank.2014}. 
Our dynamical model can robustly constrain the contribution of DM within the radius of the outermost kinematic aperture. This constraint is directly related to $M_{200}$, assuming the validity of equation \ref{DM:c}. 

\subsubsection{Stellar Mass}
We construct the contribution of stellar mass to the gravitational potential by multiplying the galaxy intrinsic 3D density distribution with a stellar mass-to-light ratio. Obtaining the intrinsic 3D luminosity density of a barred galaxy from its 2D photometry image is not straightforward. We achieve this in three steps using the method described in \citet{Behzad.2021}: (1) We first decompose the galaxy 2D image into a disk and a bar, (2) we then apply a multi-Gaussian expansion (MGE) \cite[]{cappellari.2002} to fit the 2D surface densities of the disk and the bar separately, and (3) we deproject the disk and the bar separately, allowing for different assumptions about their internal 3D shapes. Finally, by combining the 3D densities of the disk and the bar, we derive the intrinsic 3D luminosity distribution of the entire galaxy. Further details for each step are provided in the following.



\subsubsection*{T\MakeLowercase{he photometric image decomposition}}
We employ GALFIT \cite[]{Peng.2010} to decompose the 2D surface brightness of NGC 4371 using a four-component model, which provides a good fit to $3.6$ $\mu$m image. This model includes a central compact S\'{e}rsic component, an exponential nuclear disk, a bar (S\'{e}rsic profile), and an exponential main disk. The goal of the GALFIT fitting is to obtain a parametrised model that fully matches the global surface brightness of the galaxy. The central compact component (point source) in the model is introduced solely to enhance the goodness of the GALFIT fit in the central pixels of the image, it is not necessary to be a physically defined component and is not considered as a classical bulge. The uncertainty in the image, derived from Poisson noise, is used to weight the data points during the fitting process. We combine the nuclear and main disks as the disk component and then subtract the disk component from the original image to obtain a residual bar. Note that in the following analysis, we only use the residual bar derived from the original image and do not use the GALFIT compact S'{e}rsic component or the bar's S'{e}rsic model.
 \par
\begin{figure}
	\centering	%
	\includegraphics[width=1\columnwidth]{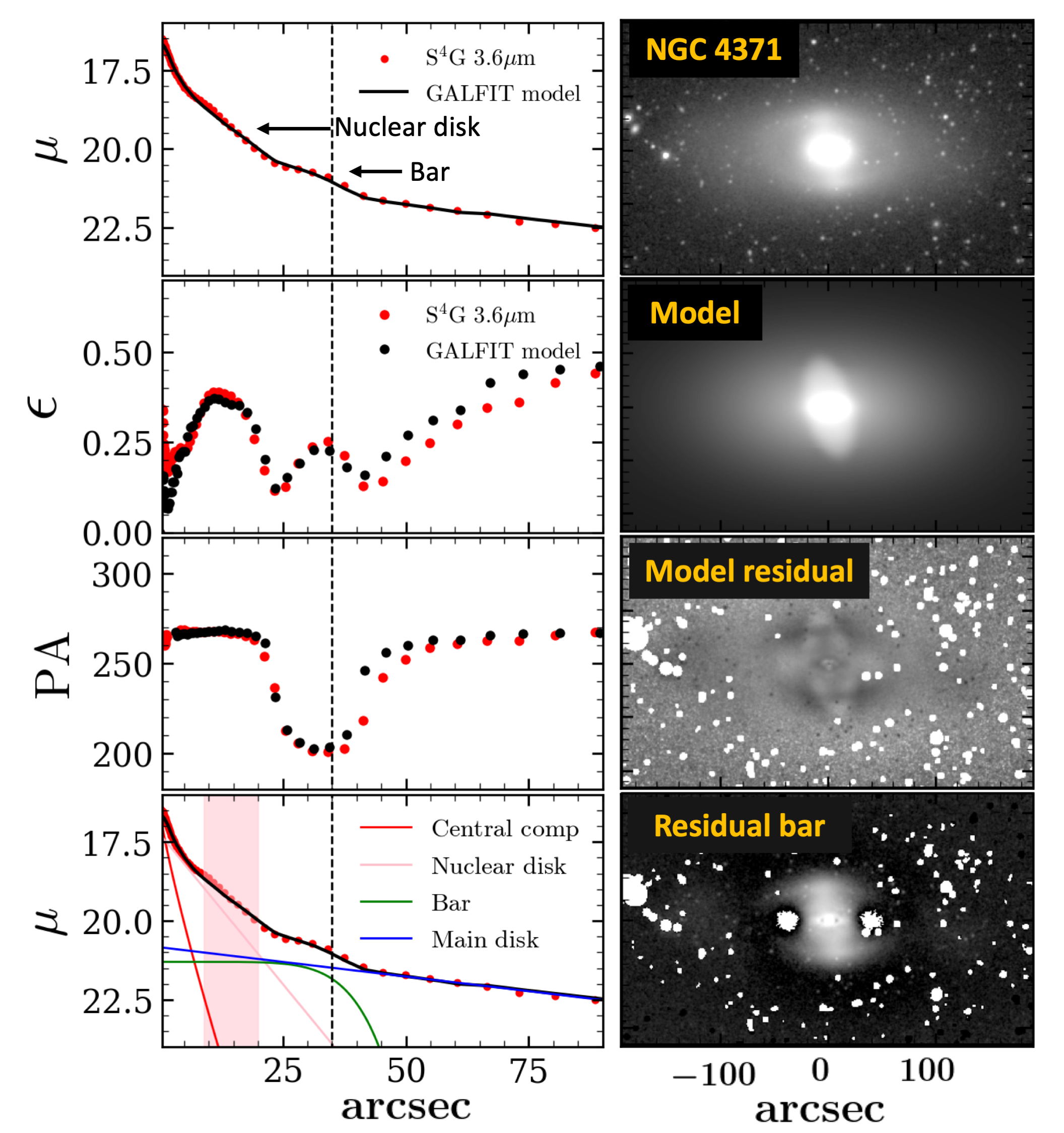}
	\caption{ The photometric analysis of the S$^4$G image and the  GALFIT best-fitting model. The left column from top to bottom shows: 1D surface brightness profile, ellipticity, and position angle along the major axis of the S$^4$G image (red) and GALFIT model (black). The last panel on the left column represents each component's 1D surface brightness profile in the GALFIT best-fitting model (central compact source, nuclear disk, bar, and main disk).  Right column from top to bottom: 2D surface brightness distribution of the NGC 4371 S$^4$G image, GALFIT best-fitting model, the residual, and the residual bar extracted by subtracting the nuclear and main disks from the original image. The black dashed line indicates the projected bar radius $R_{\rm bar}^{'} \sim 35$ arcsec. The image $x$-axis is flipped to align with the orientation of the kinematic map. The pink shaded area indicates the extent of the GALFIT nuclear disk component, from a half-light radius of  $\sim 8.8$ to $20$ arcsec.%
 }%
	\label{fig:photo}%
\end{figure}

 Table.\ref{table:galfit} presents the structural parameters derived from the fit. The right column of Fig.\ref{fig:photo} shows (from top to bottom) the S$^4$G image, the GALFIT model, the residual, and the residual bar obtained by subtracting the disk component (comprising both the nuclear and the main disks) from the S$^4$G image. The left column of Fig.\ref{fig:photo} shows (from top to bottom) the radial surface brightness profile derived from the S$^4$G image and the GALFIT model with ellipse fits to the isophotes, ellipticity, position angle, and the radial surface brightness profile of the subcomponents. From the GALFIT model, we determine the projected bar semi-major axis to be $R_{\rm bar}^{'} \sim 35$ arcsec, which aligns with the measurements reported by \cite{Gadotti.2015} using image decomposition.
 
In Fig. \ref{fig:photo} (top left), we mark the locations of the nuclear disk and the bar on the surface brightness profile derived from the S$^4$G image. 
The exponential decrease in brightness between 10-20 arcsec (1 arcsec $\sim 82$ pc) suggests the presence of a nuclear disk. The characteristic bump at a radius of $\sim 35$ arcsec corresponds to the projected bar radius we measured.

As discussed in \cite{Gadotti.2015}, the three local maxima in the ellipticity profile correspond to distinct structures of the nuclear disk, the bar, and the main disk. The nuclear disk and the main disk exhibit higher ellipticity than the bar. The two minima in the ellipticity profile signify the transition points: the first from the nuclear disk to the bar and the second from the bar to the main disk.
In the position angle (PA) profile, the nuclear disk exhibits a similar PA to that of the main disk, as also evidenced by their kinematics \cite[]{Gadotti.2015}. The sharp drop in PA at the bar radius is attributed to the bar in NGC 4371 being nearly perpendicular to the nuclear and main disks.
 
The match between the ellipticity and PA profiles of our GALFIT model and the image indicates that we have effectively captured the main features of the image. The only purpose of this photometric decomposition is to separate the disk and the residual bar component, ensuring that their combination faithfully represents the image of the entire galaxy. This will be employed to construct the 3D stellar mass density and compute the stellar gravitational potential, as detailed below. We will not rely on this photometric decomposition to study the intrinsic properties of each component; instead, we will perform a structural decomposition based on the 3D dynamical models superposed by stellar orbits at the end.
 
 \par
 
\begin{table}
	\centering
	\footnotesize
	\begin{tabular}{p{0.50\linewidth}p{0.12\linewidth}p{0.15\linewidth}}
		\hline
            \hline
              Central concentrated component (S\'{e}rsic)    \\
		\hline
            Normalized flux      & $\Sigma_{0}$    & 15.24   \\ 
            Effective radius      & $R_{e}$    & 1   \\
            S\'{e}rsic index      & $n$    & 1.10   \\
            Ellipticity           & $\epsilon$    & 0.70   \\
            Position angle     & $\rm PA$    & 90.0   \\
            \hline
            Nuclear disk   (exponential)      \\
             \hline
            Normalized flux      & $\Sigma_{0}$    & 13.47   \\ 
            Scale length      & $R_{s}$    & 5.21   \\
            Ellipticity           & $\epsilon$    & 0.45   \\
            Position angle     & $\rm PA$    & 89.0   \\
            \hline
             Bar  (S\'{e}rsic)  \\
            \hline
            Normalized flux      & $\Sigma_{0}$    & 13.93  \\ 
            Effective radius      & $R_{e}$    & 27.39   \\
            S\'{e}rsic index      & $n$    & 0.20   \\
            Ellipticity           & $\epsilon$    & 0.49   \\
            Position angle     & $\rm PA$    & 14   \\
            \hline
            Main disk   (exponential) \\
            \hline
            Normalized flux      & $\Sigma_{0}$    & 15.24   \\ 
            Scale length      & $R_{s}$    & 43.33   \\
            Ellipticity           & $\epsilon$    & 0.47  \\
            Position angle     & $\rm PA$    & 88.0   \\
            \hline
		\hline
	\end{tabular}
	\\
	\parbox{\columnwidth}{\caption{Best-fit parameters from GALFIT decomposition. Luminosity parameters are normalized using the standard method provided by GALFIT. Spatial measurements are in units of arcsec. Position angles (PA) are in degrees counter-clockwise from the image $y^{'}$-axis.}
 \label{table:galfit}}
\end{table}

\subsubsection*{MGE \MakeLowercase{Fitting}}
We combine the nuclear disk and the main disk as an axisymmetric disk component and consider the residual bar as a triaxial component. We fit MGEs to each component separately. Note that we use the residual barred bugle, which allows us to capture the triaxiality better than the fitted elliptical S\'{e}rsic bar. 

We obtain parameters $(L_{j}, q'_{j} ,\sigma '_{j}, \Delta \psi_{j}^{\prime} ) $ of the 2D Gaussians from the fitting, where $L_{j}$ is the total luminosity, $q_{j}^{\prime}$ is the projected flattening, and $\sigma_{j}^{\prime}$ is the scale length along the projected major axis of each Gaussian component $j=1 \dots N$. $ \Delta \psi_{j}^{\prime}$ is the isophotal twist of each Gaussian. The MGE fitting parameters of the barred bugle (Gaussians with $ \Delta \psi_{j}^{\prime} \neq 0 $) and the disk (Gaussians with $ \Delta \psi_{j}^{\prime} = 0 $) are presented in Table \ref{table:barmge}. 
\par

\subsubsection*{D\MakeLowercase{eprojection}}
The orientation of a projected system is defined by three viewing angles $(\theta,\varphi,\psi)$, $ \theta $ and $ \varphi $ indicate the orientation of the line-of-sight with respect to the principal axes of the object. For example, projections along the intrinsic major, intermediate, and minor axes correspond to $(\theta=90^{\circ},\varphi=0^{\circ}) $, $(\theta=90^{\circ},\varphi=90^{\circ}) $ and ($\theta=0^{\circ}$, $\varphi$ irrelevant), respectively. $ \psi $ is the position angle, which indicates the rotation of the object around the line of sight in the sky plane (see Fig. 2 in \cite{Zeeuw.1989}). The intrinsic parameters describing a 3D Gaussian component $ (\sigma_{j}, p_{j}, q_{j}) $ can be derived analytically using a set of viewing angles $ (\theta, \varphi, \psi) $ and parameters measured for the 2D Gaussians (see Eqs. (7-9) in \texttt{VdB08}).
For a rigid body comprised of multiple Gaussian components, all Gaussians are fixed to have the same viewing angles; the allowed orientations are thus the intersection of allowed viewing angles $(\theta, \varphi, \psi) $ of all the Gaussians. 

We consider the disk and the bar as two rigid body components, and we deproject them separately. Thus, we have three viewing angles $(\theta_{\rm disk}, \varphi_{\rm disk}, \psi_{\rm disk}) $ for the disk and three viewing angles $(\theta_{\rm bar}, \varphi_{\rm bar}, \psi_{\rm bar}) $ for the bar.
 \par
The disk is considered as an axisymmetric oblate system with the major axis aligned with the $ x^{\prime} $ axis of the image so that $\psi_{\mathrm{disk}} = 90^{\circ}$, and all Gaussians have $ \Delta \psi_{j}^{\prime} = 0 $, while $\varphi _{\rm disk}$ is irrelevant. The inclination angle of the disk $\theta_{\rm disk}$  is left as a free parameter, with its lower limit constrained by $\cos( \theta_{\rm disk} )^2< q'^2_{\rm min} $ where $q'_{\rm min}$ indicate the flattest Gaussian of MGEs fitted to the disk. 

The bar is triaxial, so its Gaussians can have different isophotal twists $\Delta \psi_{j}^{\prime}$. The twists of bar Gaussians are measured with respect to the major axis of the disk in the observational plane. We therefore have $ \Delta \psi_{j}^{\prime}  =\psi_{j}^{\prime}-\psi_{\mathrm{disk}} $, and $\psi_{\rm disk} = 90^{\circ}$. The intrinsic position angle $ \psi $ of an isolated triaxial system is in principle unknown \citep{bosch.2008}. However, we have a reference disk. We fix the reference bar position angle to $\psi_{\rm bar} = \psi_{\mathrm{disk}}=90^{\circ}$ and use $\Delta \psi_{j}^{\prime}$ to include the real information of the bar position angle.


Deprojection of the model is inherently not unique; there is still a wide range of possible viewing angles. We impose certain constraints to further reduce the degree of this degeneracy: 1) We assume the bar major axis aligns with the disk plane, implying the bar inclination angle $\theta_{\rm bar}$, matches the disk $\theta_{\rm disk}$. This assumption narrows down the allowed inclination angle to be $\theta_{\rm bar} = \theta_{\rm disk}$, and the angle $\varphi_{\rm bar}$ is left free. 2) We further refine the inclination angle constraint using the observationally derived inclination of $\sim 60^{\circ}$ from \cite{Gadotti.2015}, stating that $|\theta - 60^{\circ}| \leq 10^{\circ}$. These constraints significantly mitigate the degeneracy of viewing angles and narrow the range of permissible viewing angles
(see Fig. 4 in \cite{Behzad.2021}); they also lead to reasonable intrinsic shapes for barred galaxies according to our test with mock data \citep{Behzad.2021}. We acknowledge that the deprojection process does not achieve absolute uniqueness. The remaining free viewing angles, although currently undetermined, will be constrained by the following dynamical model that fits the kinematic data. This approach underscores the fact that achieving complete uniqueness in the deprojection may not be necessary for our analysis, given that the fitting to kinematic data is designed to pinpoint the most plausible viewing angles for the galaxy.


We therefore have two viewing angles as free parameters: $\theta_{\rm disk}$ and $\varphi_{\rm bar}$, which will be just denoted as $\theta$ and $\varphi$ in what follows.
\par

Once we infer the 3D luminosity density distribution with a set of viewing angles, we multiply it by a constant stellar mass-to-light ratio $M_*/L$ to obtain the 3D stellar mass distribution, which is another free parameter in the mass model. The assumption of a constant $M_*/L$ across the entire galaxy is fairly adequate for this case study. \cite{Gadotti.2015} showed that the stellar population is uniformly old across the MUSE-TIMER field. Furthermore, we utilise the S$^4$G image at 3.6 um, which traces the old stellar populations that dominate the mass budget of galaxies; the $M_*/L$ at 3.6 um should exhibit minimal variation across different populations \citep{Meidt.2012,Meidt.2014,Querejeta.2015}.

We assumed a stationary gravitational potential in the rotating frame for a barred galaxy with a triaxial bar model. Thus, the bar pattern speed $\Omega_{p}$ is left as another free parameter. 

In summary, we have five so-called free hyperparameters in the model of gravitational potential: DM virial mass $M_{\rm 200}$, inclination angle $\theta$, bar azimuthal angle $\varphi$, stellar mass-to-light ratio $M_*/L$, and the bar pattern speed $\Omega_{p}$.

 \begin{figure}
	\centering	%
	\includegraphics[width=\columnwidth]{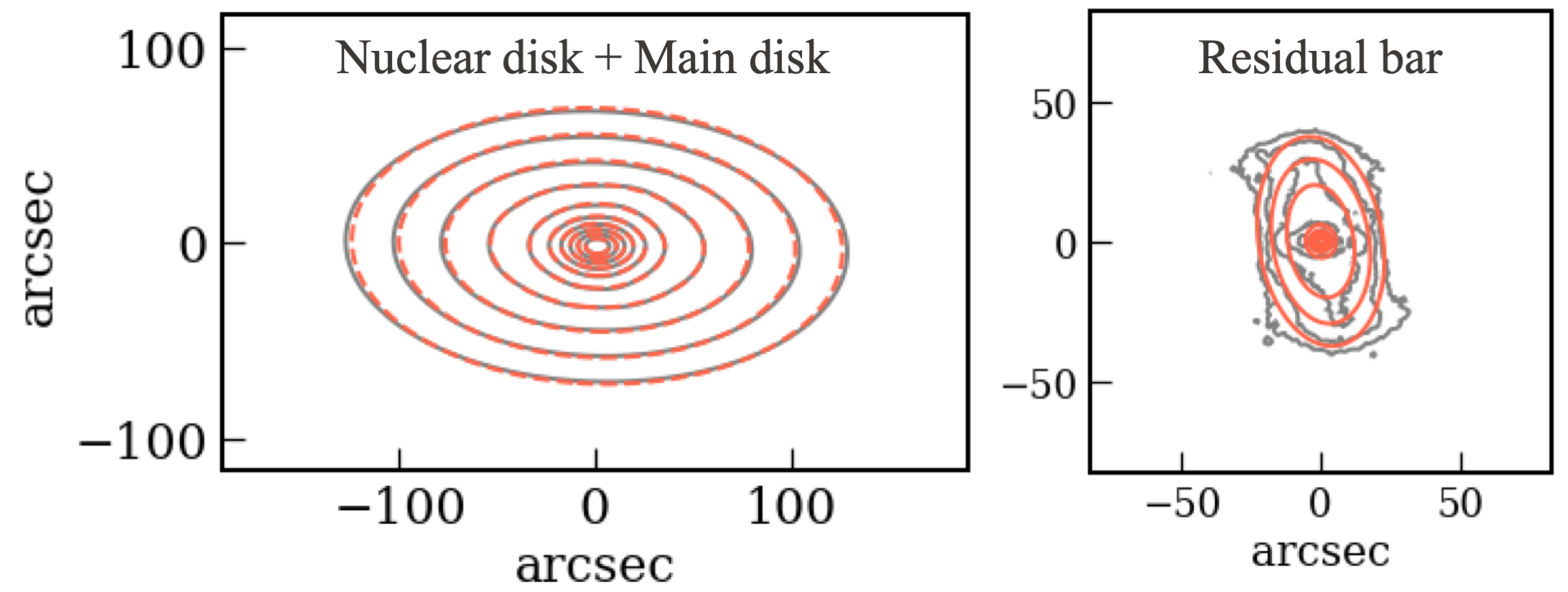}
	\caption{Left panel: The surface density contours of disk component (grey lines), overplotted with contours of the best-fitting MGE model (orange lines). Right panel: similarly for the bar, where the twist is allowed in the MGE model.}%
	\label{fig:MGE}%
\end{figure}

 \begin{table}
	\centering
	\footnotesize
	\begin{tabular}{p{0.15\linewidth}p{0.15\linewidth}p{0.15\linewidth}p{0.15\linewidth}p{0.15\linewidth}}
		\hline
		$ j $    &  $ L_{j}$ $(\mathrm{L_{\odot}pc^{-2}}) $ &  $ \sigma^{\prime}_{j} $ $ (\mathrm{arcsec}) $    & $ q^{\prime}_{j} $   &  $ \Delta \psi_{j}^{\prime} $ $(^{\circ})$   \\
		\hline
        $1$ & 4519.857 & 0.294  & 0.99 &  -59.0 \\
        $2$ & 10253.941  & 1.069  & 0.99  & -59.0     \\ 
        $3$ & 3263.905 & 2.119  & 0.981 & -66.0     \\
        $4$ & 535.953  & 18.296 & 0.58  & -63.831  \\
        $5$ & 53.586   & 45.0   & 0.614 & -59.611  \\
        \hline
        $6$ & 3842.074 & 2.343  & 0.575 & 0.0  \\
        $7$ & 4078.952 & 5.112  & 0.58  & 0.0  \\
        $8$ & 2506.333 & 8.667  & 0.57  & 0.0  \\
        $9$ & 179.587  & 11.512 & 0.965 & 0.0  \\
        $10$ & 614.981  & 14.211 & 0.57 & 0.0  \\
        $11$ & 217.091  & 31.804 & 0.57 & 0.0  \\
        $12$ & 151.360  & 60.354 & 0.57 & 0.0  \\
        $13$ & 31.390   & 106.066 & 0.57 & 0.0  \\
		\hline
	\end{tabular}
	\\
	\parbox{\columnwidth}{\caption{The MGE fitting parameters of the bar (Gaussians with $ \Delta \psi_{j}^{\prime} \neq 0 $ ) and the disk (Gaussians with $ \Delta \psi_{j}^{\prime} = 0 $ ). }
		\label{table:barmge}}
\end{table}

\subsection{Generating the Orbit Library}

We sample the initial conditions of orbits in the $ x-z $ plane, using the properties of separable models as described in \texttt{VdB08}.  In such models, tube orbits (excluding shell orbits where the outer and inner radial turning points coincide) intersect the $ x-z $ plane perpendicularly twice above $z>0$. Therefore, it is not necessary to sample the entire $ x-z $ plane. We determine the orbital energy $ E $ in a stationary frame using a logarithmic grid in radius; each energy corresponds to a grid radius $ r_{i} $ calculated by evaluating the potential at $ (x, y, z) = (r_{i}, 0, 0)$. For each energy, the starting point $ (x, z) $  is chosen from a linear open polar grid of $ (R, \phi) $, between the location of the shell orbits and the equipotential surface with zero velocity for that energy, where $R=\sqrt{x^2 + z^2}$ and $\phi=\arctan(x/z)$ (as shown by the grey area in Fig. 2 of \texttt{VdB08}). The initial starting points are sampled in the inertial frame with $ v_{y}=\sqrt{2[E-\Phi(x, 0, z)]}$, and converted to velocities in the rotating frame for the orbit integration as discussed in \cite{behzad.2022}. We sample two sets of tube orbit libraries in $ x-z $ plane, one with $v_{y}>0$ and the other with $v_{y}<0$ as these will represent different orbits when integrated with the presence of a rotating bar. A large number of starting points are sampled across the three integrals with $(n_{E} \times n_{R} \times n_{\phi}) = (30 \times 15 \times 13)$, and we adopt the dithering number to be 3 to impose the smoothness of orbit-superposition models. Hence, each orbital bundle contains $27$ orbits with close starting points.

\subsection{Weighing the Orbits} \label{S:weighing_orbits}
The constraints for the model include 1- kinematic maps, which typically encompass the observational velocity $V_o^l$ ($o$ stands for observation) and dispersion $\sigma_o^l$ for each aperture $l$, along with the Gaussian-Hermite (GH) coefficients $h_{3,o}^{l}$ and $h_{4,o}^{l}$, 2- the surface brightness in the 2D observational plane and the 3D luminosity distribution, which is deprojected from the 2D image. Note that $V_o^l$ and $\sigma_o^l$ are the parameters of the GH function obtained by the full spectrum fitting; they are not the mean velocity and its dispersion unless all higher-order moments are zero.

The model comprises a superposition of thousands of orbit bundles, where each orbit bundle, denoted as $k$, is weighted by $w_{k}$. We minimize the $\chi^2$ between the data and model to get the solution of the orbit weights. The $\chi^2$ is contributed by two parts, the fit to the luminosity distribution and the fit to kinematic maps.
\begin{equation}\label{NNLS}
\chi^{2}_{\mathrm{NNLS}}=\chi_{\operatorname{lum}}^{2}+\chi_{\mathrm{kin}}^{2}.
\end{equation}
\par

We allow a relative error margin of $1\%$ for both 2D and 3D luminosity distribution fittings. The 2D luminosity distribution, represented as $S_l$, is stored within the observational apertures on the observational plane. For each aperture $l$, the contribution from orbit bundle $k$ is expressed as $S^{l}_{k}$. Similarly, the 3D density distribution, denoted as $\rho_n$, is catalogued in a three-dimensional grid comprising a total of 360 bins. Within this grid, the contribution of each orbit bundle $k$ to a specific bin $n$ is indicated by $\rho^{n}_{k}$. We thus have
\begin{equation}
\begin{split}
\chi_{\operatorname{lum}}^{2} &=\chi_{\mathrm{S}}^{2}+\chi_{\rho}^{2} \\
&= \sum_{l=1}^{ N_{\rm kin}}\left[\frac{\sum_{k} w_{k} S^{l}_{k}-S_{l}}{0.01 S_{l}}\right]^{2} +
\sum_{n=1}^{360}\left[\frac{\sum_{k} w_{k} \rho^{n}_{k}-\rho_{n}}{0.01 \rho_{n}}\right]^{2},
\end{split}
\end{equation}
where $N_{\rm kin}$ represents the total number of apertures in a single kinematic map, while $w_k$ denotes the weight assigned to orbit $k$. 
\par

From observations, we describe the LOSVD profile $f_l$ in each aperture $l$ as a GH distribution \cite[]{Gerhard.1993, Marel.1993} with parameters $(V_o^l, \sigma_o^l, h_{3,o}^l, h_{4,o}^l)$ and corresponding errors $(\Delta V_o^l, \Delta \sigma_o^l, \Delta h_{3,o}^l, \Delta h_{4,o}^l)$. When $V_o^l$ and $\sigma_o^l$ are chosen as the centre and the width of the best-fitting Gaussian approximating the original LOSVD, this resulted in $h_{1,o}^l= h_{2,o}^l=0$.

We denote the LOVSD contributions of orbit bundle $k$ at aperture $l$ as $f_k^l$. If we expand $f^l_k$ in a GH series also with the central velocity and dispersion fixed at the observed $V_o^l$ and $\sigma_o^l$, then the resulting GH coefficients $h_{n, k}^{l}$ with $n=1,2,3,$ and $4$ will contribute linearly to the observations, so that
\begin{equation}
\chi_{\mathrm{kin}}^{2}=\sum_{l=1}^{ N_{\rm kin}} \sum_{n=1}^{n_{\rm GH}}\left[\frac{\sum_{k} w_{k} S_{k}^{l} h_{n, k}^{l} - S_{l} h_{n, o}^l}{S_{l} \Delta h_{n, o}^l}\right]^{2},
\end{equation}
where the model predictions are luminosity weighted in the same manner as the observations, and $n_{\rm GH}$, which represents the number of kinematic moments used for the fitting, is set to 4 here. The errors of $(\Delta V_o^l, \Delta \sigma_o^l, \Delta h_{3,o}^l, \Delta h_{4,o}^l)$ are usually provided directly from observations, while we derive $\Delta h_{1,o}^l, \Delta h_{2,o}^l$ following \citet{Rix.1997}. The luminosity density is usually easy to fit, so that $\chi_{\mathrm{kin}}^{2}$ is the dominant term contributing to goodness of fit $\chi^{2}_{\mathrm{NNLS}}$ \cite[e.g.,][]{ling.1018}. 
We use the nonnegative least squares (NNLS) implementation \cite[]{Lawson.1974} to find the solution of orbit weights by minimizing the $\chi^2_{\mathrm{NNLS}}$ between data and model following \texttt{VdB08}.

\subsection{Exploring the Parameter Space}

We have five free hyperparameters in the model: $1$- the stellar mass-to-light ratio $M_{*}/L$ at 3.6 $\mu$m band,  $2$- the inclination angle $\theta$, $3$- the bar azimuthal angle $\varphi$, $4$- the bar pattern speed $\Omega_{\rm p}$, and $5$- the DM virial mass $M_{200}/M_{*}$.

In our search for the best-fitting model, we employ an optimized grid iterative process.  We start with initial guesses of the hyperparameters. We then walk two steps in each direction of the parameter grid by taking relatively large intervals of $ 0.2 $, $ 6 $, $ 2 $, $ 2 $, and $ 0.4 $ for $ M_{*}/L $, $ \Omega_{\rm p} $, $ \theta $, $ \varphi $, and $\mathrm{log_{10}}(M_{200}/M_{*})$, respectively. 

After completing the initial models, we select those with a $\chi^{2} - \chi^{2}_{\rm min} < 100 \times \sqrt{2 n_{\rm GH} N_{\rm kin}}$ from the existing models. The factor of $100$ is an empirical choice. Then, we run new models around the selected models. This iterative process continues until we identify the model with the lowest $\chi^{2}$, ensuring that all models in its vicinity are also evaluated. Then, we halve the parameter step sizes to better sample the grids around the best-fitting models.
 
Finally, once again we repeat the iterative search process, but this time with an increased threshold of $\chi^{2} - \chi^{2}_{\rm min} < 500 \times \sqrt{2 n_{\rm GH} N_{\rm kin}}$. This approach is adopted to prevent the process from getting trapped in a local minimum and to guarantee that all models within a $3\sigma$ confidence level are thoroughly calculated.
\par 
In classical statistic analysis for analytic models fitting to data, the $1\sigma$ confidence level is determined by $\Delta \chi^{2}=1$ for one degree of freedom. However, it is unsuitable for our case where the model numerical noise dominates the $\chi^2$ \citep{Lipka2021}. In \cite{ling.1018},  $\Delta \chi^{2} \equiv \sqrt{2 n_{\rm GH} N_{\rm kin}}$ is adopted as $ 1 \sigma $ confidence level, which is consistent with the $\chi^2$ fluctuation caused by numerical noise of their models for CALIFA galaxies.

Here, we adopt a similar approach to calculate the $1\sigma$ confidence level, utilising a bootstrapping process. This involves randomly perturbing the kinematic data within its error margins and generating $100$ new kinematic maps. Subsequently, we re-fit only our best-fitting model—which has a fixed potential and orbit library—to these $100$ perturbed data sets. The standard deviation of the $\chi^2$ values from these fittings is then used to define the $\chi^2$ fluctuation attributable to the model's numerical noise, which is related to various factors such as the non-uniqueness of the orbit weight distribution, degree of freedom, observational errors, etc. Through this method, we determined that the $1\sigma$ confidence level for models constrained by TIMER data is approximately $\sim 0.5 \times \sqrt{2 n_{\rm GH} N_{\rm kin}}$, and for those constrained by ATLAS3D data, it is about $\sim 2 \times \sqrt{2 n_{\rm GH} N_{\rm kin}}$.


\section{Results}\label{S:results}
By exploring the parameter space, we obtain the best-fitting models that match both the target density distribution and all kinematic features. We anticipate achieving robust constraints on the free parameters in the gravitational potential (Section 4.1). This will enable us to determine the bar pattern speed (Section 4.2), the enclosed mass profiles (Section 4.3), and the viewing angles, thus the 3D intrinsic structure of the galaxy. We will facilitate a dynamical structure decomposition based on the 3D model, allowing us to explore the intrinsic properties of each structural component (Section 4.4).

\subsection{Best-fitting Models}
We consider all models that fall within the $1\sigma$ confidence level and compute the mean and standard deviation of each parameter for these models. These calculations are then used to determine the best-fitting parameter and its corresponding $1\sigma$ error.

From the models constrained by TIMER, we obtained the best-fit parameters as $\theta = 60 \pm 2 ^\circ$, $\varphi = -12 \pm 1 ^\circ$, $\Omega_{\rm p} =  23.6 \pm 2.8 \hspace{.08cm} \mathrm{km \hspace{.04cm} s^{-1} \hspace{.04cm} kpc^{-1} }$, $M_{*}/L_{3.6 \mu m}=1.05\pm 0.05$ $M_{\odot}/L_{\odot ,3.6 \mu m}$, and DM virial mass $\log_{10}(M_{200}/M_{*})=2.4 \pm 0.5$.

From the models constrained by ATLAS3D, we obtain $\theta = 60 \pm 2 ^\circ$, $\varphi = -12 \pm 1 ^\circ$, $\Omega_{\rm p} =  22.4 \pm 3.5 \hspace{.08cm} \mathrm{km \hspace{.04cm} s^{-1} \hspace{.04cm} kpc^{-1} }$, $M_{*}/L_{3.6 \mu m}=0.99\pm 0.11$ $M_{\odot}/L_{\odot ,3.6 \mu m}$, and DM virial mass $\log_{10}(M_{200}/M_{*})=3.0 \pm 0.6$. The parameters obtained from the two sets of models are generally consistent with each other. There is a larger uncertainty in the model constrained by ATLAS3D data, due to the smaller spatial coverage of ATLAS3D compared to the MUSE data. The parameters grid of all models is included in Appendix Fig.~\ref{fig:muse_bar_chi2} and Fig.~\ref{fig:atlas_bar_chi2}. 

In $3.6$ $\mu$m band, by assuming a Chabrier initial mass function (IMF), the stellar population synthesis gives an average of $\Upsilon_{3.6 \mu m, *}^{\mathrm{Chab}} \sim 0.6$, with uncertainty of 0.1 dex \cite[]{Meidt.2014}, which is lower than the dynamical stellar mass-to-light ratio $\Upsilon_{3.6 \mu m, *}^{\mathrm{dyn}} \sim 1.05$ we obtained. If we scale it to a Salpeter IMF multiplied by a factor of 1.8, then $\Upsilon_{3.6 \mu m, *} ^{\mathrm{Salp}} \sim 1.08$ is consistent with our dynamical results within $1\sigma$ uncertainty.

\par
\begin{figure*}
	\centering	%
	\includegraphics[width=2.1\columnwidth]{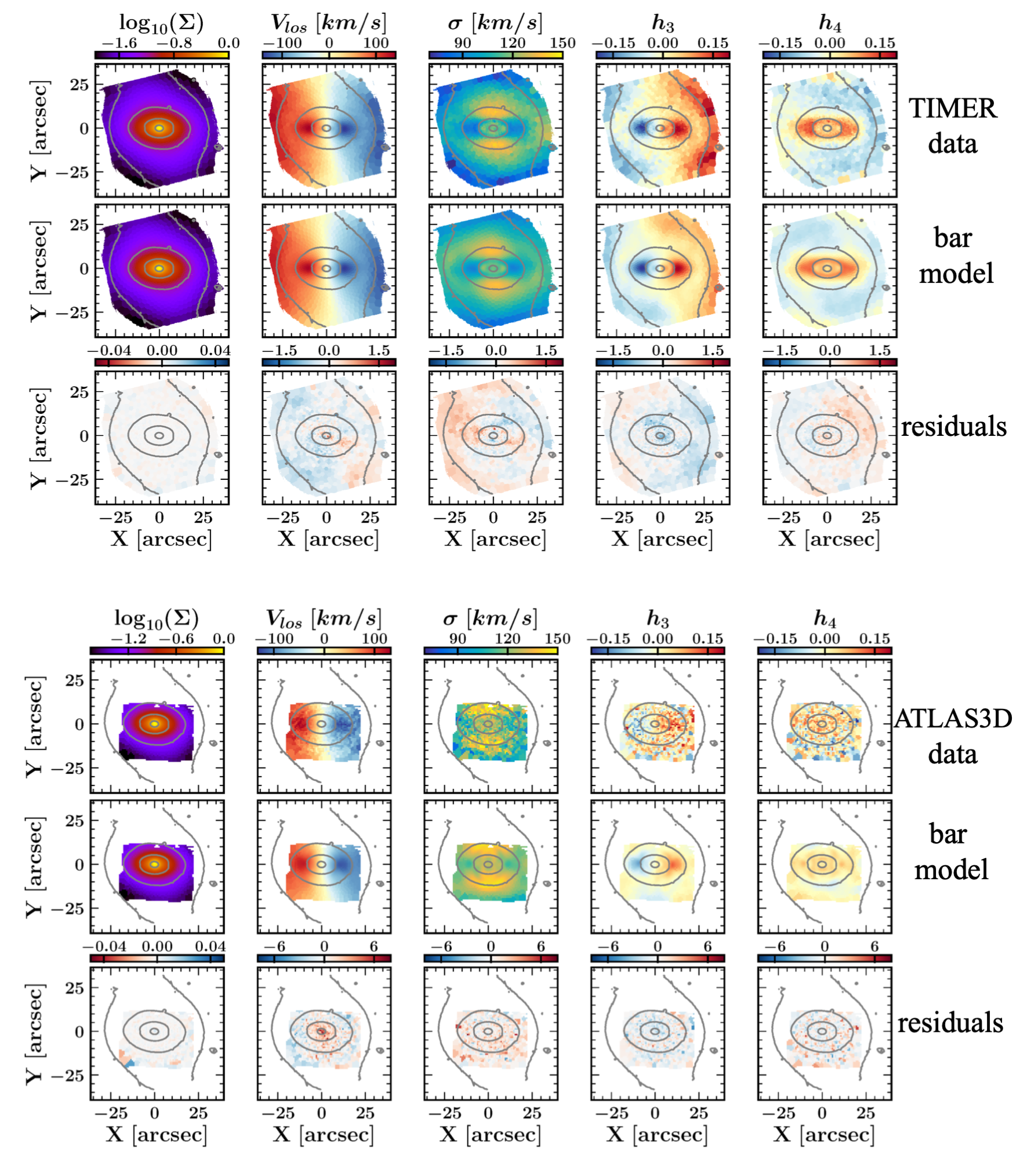}
	\caption{The best-fitting models of NGC 4371 to the data from TIMER (top panel) and ATLAS3D (the bottom), respectively. In each panel, columns from left to right represent the 2D surface density, velocity, velocity dispersion, $h_{3}$, and $h_{4}$, rows from top to bottom are the observational data, the best-fit Schwarzschild barred model, and the residuals. The overplotted grey contours are the surface brightness of the NGC 4371. The TIMER data covers the regions with $r\lesssim 35$ arcsec, while ATLAS3d covers only the inner regions with $r \lesssim 25$ arcsec. The models are well-matched with both sets of data. }%
	\label{fig:kinematic_bar}%
\end{figure*}

We present the best-fitting models of NGC 4371 in Fig. \ref{fig:kinematic_bar}. Columns from left to right displaying the 2D surface density, LOS velocity, velocity dispersion, $h_{3}$, and $h_{4}$.  The first three rows depict the TIMER data covering $\sim 35$ arcsec in radius, followed by the best-fit Schwarzschild bar model and the residuals (calculated as the difference between the TIMER data and the model, normalized by the uncertainties in each TIMER data bin). The fourth to sixth rows illustrate the ATLAS3D data within $\sim 20$ arcsec in radius, the corresponding best-fit model, and the residuals. All panels include grey contours indicating the surface brightness of NGC 4371. 

Our models successfully reproduce key kinematic features, ranging from the nuclear disk to the outer bar regions. They demonstrate improved fitting to the kinematic maps compared to nearly axisymmetric models that do not explicitly include the bar. In appendix \ref{Appendix}, we discuss the limitations of an axisymmetric model, illustrating how it still can fit the data while significantly biasing the internal properties of the best-fit model (see Fig. \ref{fig:axi_bar}).


\subsection{The Bar Pattern Speed and Rotation Parameter}

 \begin{figure}
	\centering	%
        \includegraphics[width=\columnwidth]{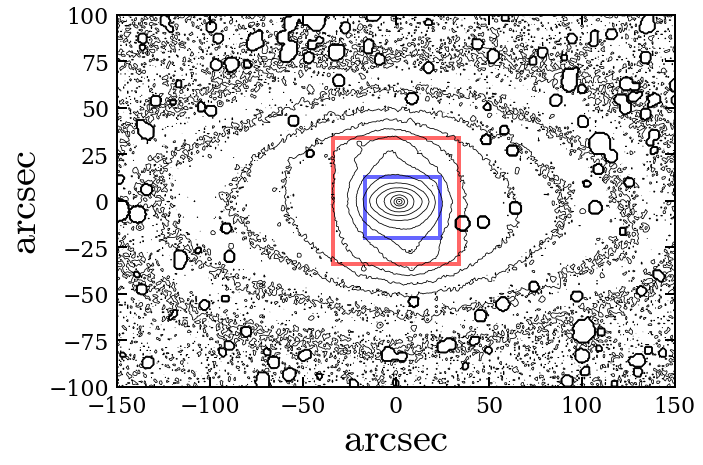}
	\includegraphics[width=\columnwidth]{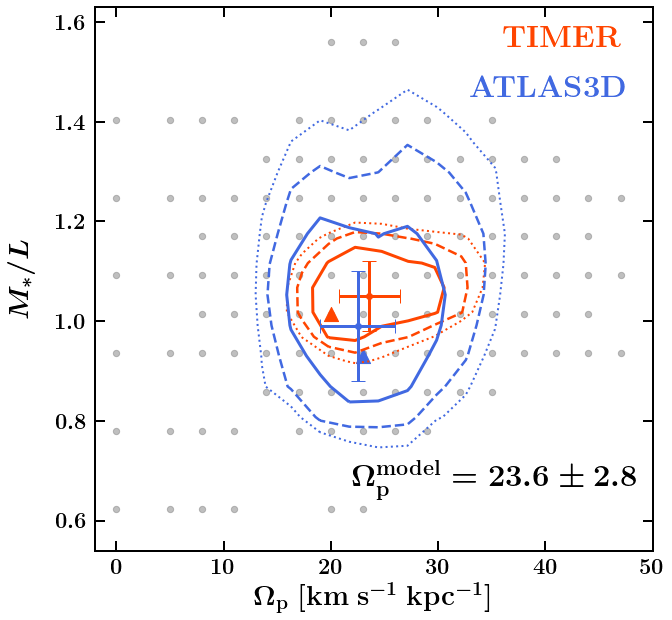}
	\caption{The bar pattern speed obtained from the two sets of models constrained by TIMER and ATLAS3D separately. Top panel: the spatial coverage of TIMER (red) and ATLAS3D (blue) data comparing to the whole galaxy, with contours showing the $3.6$ $\mu$m S$^4$G image. Bottom panel: the model parameter grid bar pattern speed $\Omega_{\rm p}$ versus the stellar mass-to-light ratio $M_{*}/L_{3.6 um}$. The contours indicate the $1 \sigma$ (solid line), $2 \sigma$ (dashed line), and $3 \sigma$ (dotted line) uncertainties for the best-fitting models using TIMER (red) and ATLAS3D (blue). The small dot with error bars indicates the mean and $1 \sigma$ uncertainty obtained from the models within the $1\sigma$ confidence level. The triangles indicate the best-fitting model for each data set. The pattern speed obtained from the two sets of models are well consistent with each other, with smaller uncertainty from the model constrained by TIMER.}%
	\label{fig:ps}%
\end{figure}

\begin{figure*}
	\centering	%
	\includegraphics[width=2\columnwidth]{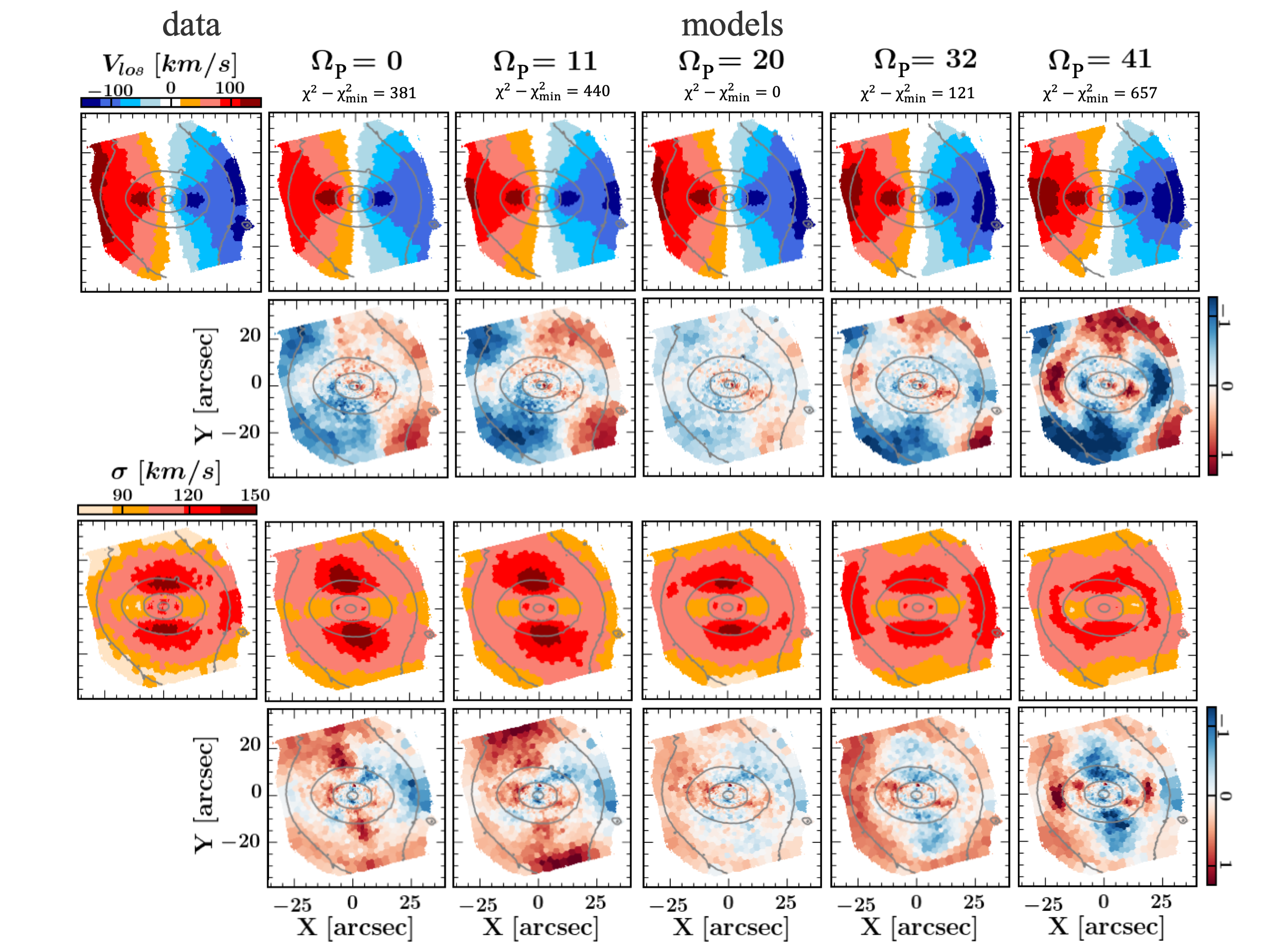}
	\caption{The kinematic maps predicted by models with different pattern speeds. The first column shows velocity (top) and velocity dispersion (bottom) from TIMER data, the following columns from left to right are the model predictions with pattern speed of  $\Omega_{\rm p}=0, 11, 20, 32, 41$ $\rm km s^{-1} kpc^{-1}$ in which $\Omega_{\rm p}=20$ $\rm km s^{-1} kpc^{-1}$ is the best-fitting model. All other parameters were kept the same as the best-fitting model. The second and fourth rows represent the corresponding residuals. The residuals of both the velocity and velocity dispersion maps become significantly larger with the pattern speed deviating from the best-fitting model ($\Omega_{\rm p}=20 $ $\rm km s^{-1} kpc^{-1}$).}  %
	\label{fig:kinematic_PS}%
\end{figure*}

 \begin{figure}
	\centering	
        \includegraphics[width=\columnwidth]{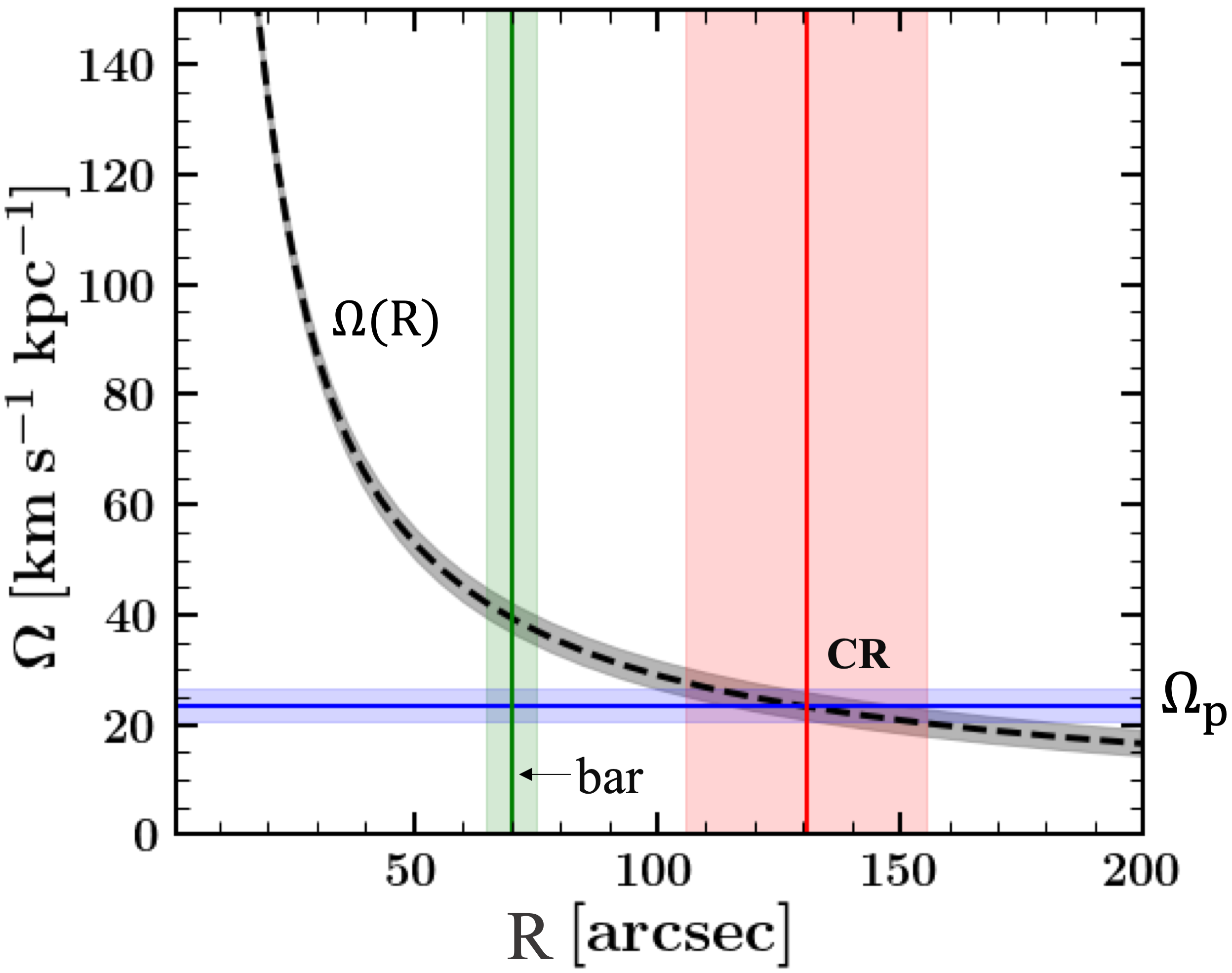}
	\caption{Analysis of corotation resonance for models within $1 \sigma $ confidence level constrained by TIMER. The black dashed curve and shaded regions indicate the local angular frequency profile $ \Omega (R)$ and its $1 \sigma $ uncertainty. The horizontal blue line with shaded regions represents the measured bar pattern speed and its $ 1 \sigma $ uncertainty. The vertical green line indicates the bar radius $R_{ \rm bar} \sim 69$ arcsec measured by photometric decomposition, with the shaded regions indicating the lower/upper limits of the bar radius. The vertical red line with shaded regions indicates the location of corotation resonance $(R_{ \rm cor})$ at $ \sim 132$ arcsec, along with its $1 \sigma $ uncertainty.}
	\label{fig:co-rot}
\end{figure}

\subsubsection{The bar pattern speed}
As depicted in the upper panel of Fig.\ref{fig:ps}, neither the TIMER nor the ATLAS3D data encompass the disk-dominated regions of NGC 4371, which complicates the determination of the bar pattern speed using the TW method. However, the bar pattern speed can be well constrained by the Schwarzschild model, even with limited data coverage. This is because the orbit-superposition method utilises full kinematic information, including higher velocity moments, which strongly depends on the internal orbital structure within the bar regions. Even when data cover only the inner region of the bar, it still contains crucial kinematic information that reflects the overall internal kinematics and orbital structure of the bar. This method has been verified through tests on a set of mock data representing various orientations and covering only the inner region of the bar \cite[]{behzad.2022} (see also other independent tests by \cite{Vasiliev.2019c} and \cite{Dattathri.2023}). These tests, which all used data that covered only the inner region of the bar, consistently report the recovery of pattern speed with a high degree of precision, with an uncertainty of less than 10\%.

In the bottom panel of Fig. \ref{fig:ps}, we illustrate the parameter space of $\Omega_{\rm p}$ versus $M_{*}/L$ constrained by the TIMER (red) and ATLAS3D (blue) data. Solid, dashed, and dotted lines indicate the confidence level of $1 \sigma$,  $2 \sigma$,  and $3 \sigma$ regions, respectively. We find the bar pattern speed of $\Omega_{\rm p} =  23.6 \pm 2.8 \hspace{.08cm} \mathrm{km \hspace{.04cm} s^{-1} \hspace{.04cm} kpc^{-1} }$ and $\Omega_{\rm p} =  22.4 \pm 3.5 \hspace{.08cm} \mathrm{km \hspace{.04cm} s^{-1} \hspace{.04cm} kpc^{-1} }$ from the models constrained by the TIMER and ATLAS3D data, respectively. The ATLAS3D data yield similar bar pattern speed but with larger uncertainties due to limited data coverage, in contrast to the TIMER data that fully encompass the bar regions. However, the ATLAS3D coverage still contains important information. It fully covers the peak velocity dispersion in the lower part, the higher values of $h_4$ in the central region, and the rotation in the nuclear disk region reflected in both velocity and $h_3$, and they are consistent with the MUSE data at the same regions. 
\par
To demonstrate the impact of pattern speed on various moments in kinematic maps, Fig. \ref{fig:kinematic_PS} compares the maps generated by models with different pattern speeds. The first column displays the velocity (top) and velocity dispersion (bottom) derived from the TIMER data. The models, with pattern speeds of $\Omega_{\rm p}=0, 11, 20, 32, 41$ $\rm km s^{-1} kpc^{-1}$, are presented by columns from left to right. Among them, $\Omega_{\rm p}=20$ $\rm km s^{-1} kpc^{-1}$ represents the best-fitting model. All other parameters are kept the same as the best-fitting model. The second and fourth rows display the residuals, calculated as the difference between the TIMER data and the model, divided by the uncertainties in the TIMER data for each bin. The residual maps clearly demonstrate that as the pattern speed deviates from $\Omega_{\rm p}=20$ $\rm km s^{-1} kpc^{-1}$ the value of $\chi^2$ increases significantly. When $\Omega_{\rm p}$ is smaller than the optimal value, the model predicts insufficient regular rotation in the outer bar regions, and the dispersion in the two high-dispersion lobes becomes too large. In contrast, a larger $\Omega_{\rm p}$ leads to predictions of an overly strong regular rotation in the outer bar regions, causing the high-dispersion lobes to vanish. Both the velocity and dispersion maps are crucial for accurately constraining the pattern speed within our Schwarzschild model.

\par
In addition to requiring extended data coverage, the TW method also demands an optimal orientation of the disk and bar for accurate pattern speed measurements. As discussed in \cite{Zou.2019}, a disk nearly face-on/edge-on or a bar nearly parallel/perpendicular to the disk major axis could lead to a wrong measurement of the pattern speed. The bar in NGC4371 is nearly perpendicular to the main axis of the disk. We attempted to apply the TW method to NGC 4371 and yielded a pattern speed of approximately $\sim 220$ $\rm km\,s^{-1} kpc^{-1}$, which is very likely to be erroneous. This is much higher than any pattern speed ever seen in observation or simulations.  


\subsubsection{The dimensionless bar rotation parameter $\mathcal{R}$}\label{section:R}
\par
For each model within the $1\sigma$ confidence level, we compute the local angular frequency $\Omega(R)$ using the model potential $\Phi$, following \cite{Binny.2008}: 
\begin{equation}
\Omega(R) = \sqrt{\frac{1}{R} \frac{\mathrm{d} \Phi}{\mathrm{~d} R}}
\end{equation}

To obtain a more accurate estimation of $\Omega(R)$ in our triaxial potential, we use quasi-axisymmetric approximation and calculate the average frequencies along the major and minor axes of the bar in the galaxy plane. However, since our model includes an axisymmetric disk, the $\Omega(R)$ along both axes are similar outside the bar.
In Fig.\ref{fig:co-rot}, we present the angular frequency (black curve) as a function of radius.
Assuming that the bar rotates as a solid body with angular velocity $\Omega_{p}$, we determine the corotation radius $R_{\rm cor}$ using $\Omega(R_{\rm cor})=\Omega_{\rm p}^{\rm model}$, where $\Omega_{\rm p}^{\rm model}$ is the pattern speed obtained by the corresponding model. At $R_{\rm cor}$, the materials within the disk rotate at the same angular velocity as the bar.
For each of our models, we infer the co-rotation radius by analyzing the angular frequency profile and the bar pattern speed, resulting in a single value for $R_{\rm cor}$ without associated uncertainty. Thus, we compute the co-rotation radius for all models within the $1\sigma$ region, which leads to a mean and standard deviation of $R_{\rm cor}=131\pm24 $ arcsec ($\sim 10.7$ $\rm kpc$).
\par

For the bar length of NGC 4371, we obtained the projected bar radius of $R_{\rm bar}^{'} = 35$ arcsec from the photometric decomposition. With the inclination angle of $\theta = 60 ^{\circ}$ and the bar angle of $\varphi = -12^{\circ}$ from our best-fitting Schwarzschild mode, we obtain the intrinsic bar radius of $R_{\rm bar} \sim 69$ arcsec ($\sim 5.7$ $\rm kpc$) using $R_{\rm bar}=R_{\rm bar}^{'} \sqrt{\sin ^2 \varphi+\cos ^2 \varphi / \cos ^2 (60^{\circ})}$ \cite[]{Gadotti.2007}. Note that $\varphi$ here differs from that in \cite{Gadotti.2007}, so the formula is modified accordingly.
We considered the range of 64 to 75 arcsec as the upper and lower limits of bar length following \cite{Erwin.2008}, who employed the IRAF ellipse package to estimate the possible maximum and minimum bar lengths. 

To estimate the dimensionless bar rotation parameter $ \mathcal{R} \equiv R_{\rm cor}/R_{\rm bar}$ and its uncertainties for NGC 4371, we first compute the co-rotation radius for each model, which yields a single value without error. We then divide the co-rotation radius by the bar length, as well as the upper and lower limits of the bar length. This results in three possible values for the bar rotation parameter for each model, including its upper and lower limits. For example, in the best-fit Schwarzschild model, the bar rotation parameter is $\mathcal{R} = 2.5 \pm 0.2 $. However, we report the bar rotation parameter for the galaxy by calculating the mean and standard deviation of bar rotation parameter values across all models within the $1\sigma$ region. This results in a bar rotation parameter of $\mathcal{R} = 1.88 \pm 0.37 $. A bar in a galaxy can be classified as fast (if $1 < \mathcal{R} < 1.4$) or slow (if $ \mathcal{R} > 1.4$) \cite[]{Debattista.2000}. Therefore, we conclude that NGC 4371 has a slow bar, even considering the lower limit of $\mathcal{R} = 1.51$, although it is near the borderline.

\subsection{Mass Profile}
\begin{figure}
	\centering	%
	\includegraphics[width=\columnwidth]{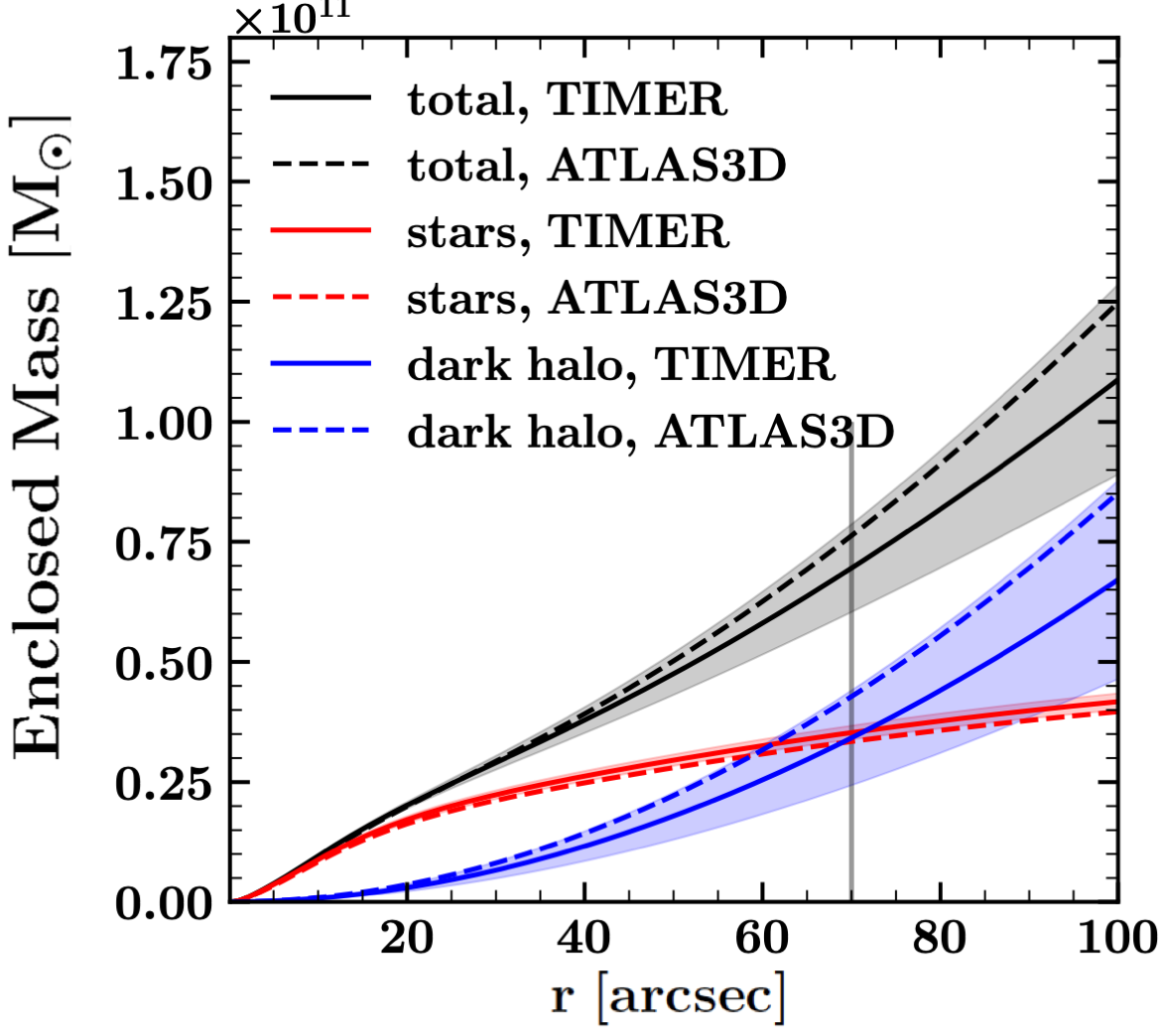}
	\caption{ 
    The enclosed mass profiles of the best-fitting bar models constrained by TIMER (solid line) and ATLAS3D (dashed line). The red, blue, and black curves represent the stellar mass, dark matter mass, and total mass, respectively. The shaded regions indicate the $1 \sigma$ uncertainty from the set of models constrained by TIMER. The vertical grey line indicates the bar length $R_{\rm bar} \sim 70$ arcsec. }%
	\label{fig:mass}%
\end{figure}

We show the enclosed mass profiles of NGC 4371 in Fig. \ref{fig:mass}. The black, red, and blue curves represent the total enclosed mass, stellar mass, and DM mass profiles, respectively. The shaded regions represent the $1 \sigma$ uncertainty from the models with TIMER. The total mass profiles are well constrained with $1 \sigma$ uncertainties of $~10$ percent within data coverage of $35$ arcsec.
The mass profiles from the models constrained by ATLAS3D are consistent with those from TIMER; however, they exhibit relatively larger uncertainties in the outer regions due to the smaller data coverage of ATLAS3D. Our model predicts that NGC 4371 has a high dark matter fraction of $ M_{\rm DM}/ M_{\rm total} \sim 0.51 \pm 0.06$ within the bar region.  
It is similar to the dark matter fraction ($\sim0.53 \pm 0.02 $) of NGC 4277 inside the bar regions, which is also a slow bar \cite[]{Buttitta.2023}. These results are aligned with cosmological simulations, suggesting that fast bars are typically found in baryon-dominated disks \cite[]{Fragkoudi.2021}.  
To confirm the correlation between the bar rotation parameter and the dark matter content in real galaxies, it is necessary to apply these measurements to a substantial number of barred galaxies.

\subsection{Dynamical Structure Decomposition and Their Internal Properties}
\subsubsection{Orbital decomposition}
After finding the best-fit Schwarzschild model and constraining the model parameters linked to the overall galaxy dynamics, we obtain the orbits and their corresponding weights that describe the galaxy kinematic and photometric data. We now aim to dynamically decompose this galaxy based on the orbital properties in the best-fit models. The dynamic decomposition is fundamentally different from the photometric decomposition described earlier. In photometric decomposition, different assumed components are combined to fit the photometric image. In contrast, dynamical decomposition does not involve any further fitting process, since the best-fit Schwarzschild model is already established. Instead, we utilise the orbits from the best-fit model and group orbits with similar kinematic and morphological features based on our arbitrary (but meaningful) criteria to construct different components. To achieve this, we first characterise each orbit using various parameters. Then, we classify the orbits into different groups based on their properties and rebuild the density and kinematic maps with each group of orbits. This process allows us to better understand the contributions of different orbital families to the overall structure and dynamics of the galaxy.

\begin{figure*}
	\centering	%
	\includegraphics[width=2\columnwidth]{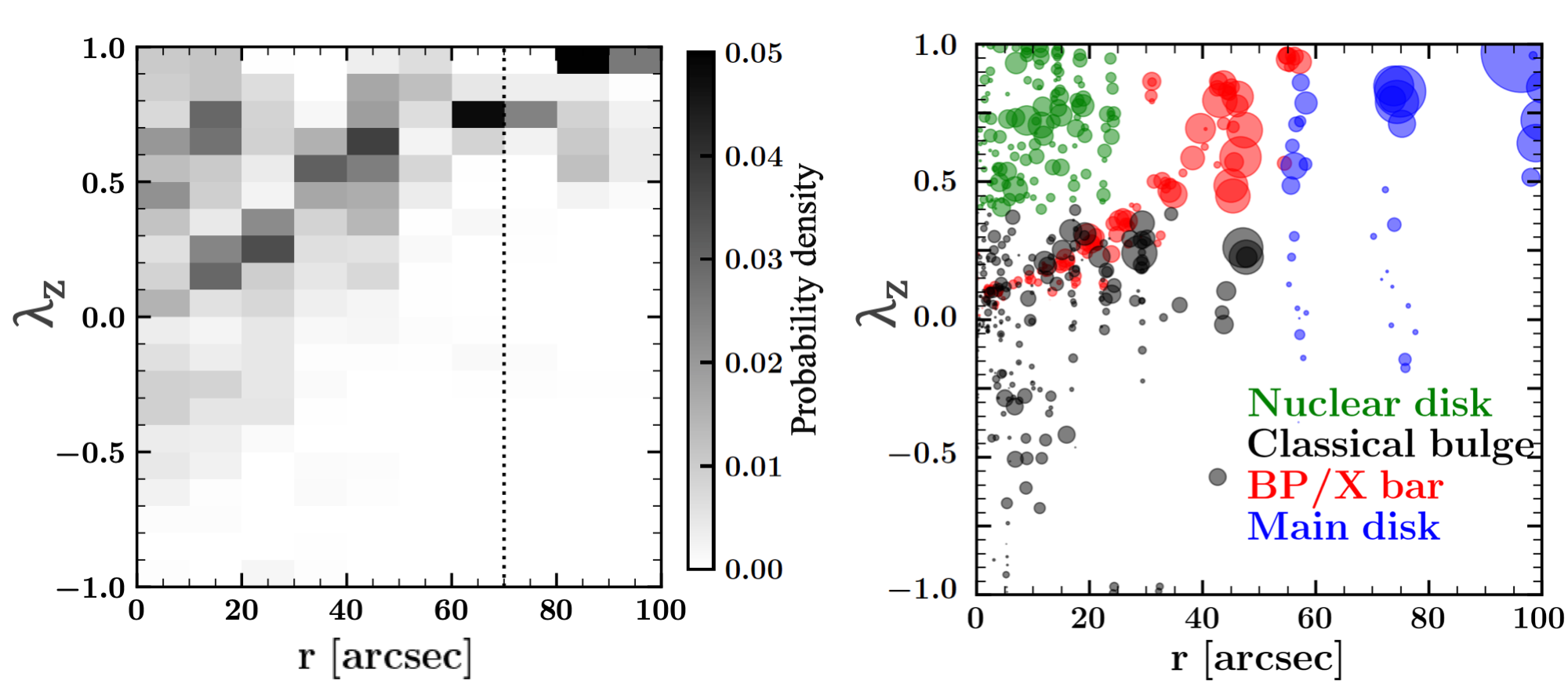}
	\caption{The stellar orbit distribution of the best-fitting model constrained by TIMER data. {\bf Left panel}: the probability density distribution of orbits in the space of circularity $\lambda_{z}$ vs. time-averaged radius $r$. Darker colours indicate a higher probability density, as the colour bar indicates. The vertical dotted lines indicate the bar radius and kinematics data extent of TIMER. {\bf Right panel:} the location of orbits making up different structures: BP/X bulge (red), classical bulge (black), nuclear disk (green), and main disk (blue). The larger symbol size corresponds to orbits with higher weights from the minimum value of $10^{-7}$ to a maximum of $10^{-1}$ (the total weight of all orbits is $= 1$).} %
	\label{fig:cirty}%
\end{figure*}

We first characterize the stellar orbits by two parameters following \cite{ling.1018}: the circularity $\lambda_{z}$ defined as the angular momentum $L_z$ (recorded in an inertial frame) normalized by the maximum angular momentum allowed by a circular orbit $L_{c}$ with the same binding energy. The radius $r$ is taken as the average of particles along an orbit stored with equal time steps in the Schwartzchild model. We show the stellar orbit distribution of the best-fitting model constrained by TIMER in the space of circularity $\lambda_z$ versus radius $r$ in the left panel of Fig. \ref{fig:cirty}. The vertical dotted line indicates the bar radius, which is within the kinematic coverage of TIMER data. Prograde circular orbits have $\lambda_z \sim 1$, box orbits with no net rotation have $\lambda_z \sim 0$, and circular retrograde orbits have $\lambda_z \sim -1$.  The internal stellar orbit distribution should be well constrained by the Schwarzschild model, as demonstrated in previous tests with many mock galaxies \citep{ling.1018, Zhu2018b, Zhu.2022, Jin.2020}


\par

To further identify the orbits that make up the bar, we employ the frequency analysis of the orbits in the best-fitting model using the “Numerical Analysis of Fundamental Frequencies (NAFF)" software \footnote{\url{https://bitbucket.org/cjantonelli/naffrepo/src/master/}}. We compute orbital frequencies in both Cartesian and cylindrical coordinates and classify the orbits into various types following the criteria established by \cite{Valluri.1998, Valluri.2016}. This orbit classification is crucial for identifying different structures within the galaxy, particularly those associated with bars.
\par  

We begin by segregating all orbits within the bar, considering those with apocenter radii smaller than the bar radius ($R_{\rm apo}<R_{\rm bar}$). These orbits are then classified into three groups: BP/X bar, Classical bulge, and Nuclear disk. The remaining orbits with $R_{\rm apo}>R_{\rm bar}$ are considered to form the main disk. Detailed descriptions of the orbits in each category are as follows:

\textbf{(1) BP/X bar}: including $x_{1}$, banana ($1:2$ resonance), periodic and non-periodic $z$-tube orbits which are within the bar and elongated along the bar. These orbits exhibit prograde motion and display an X-shaped structure in edge-on projected surface density, consistent with previous studies \cite[]{Portail.2015, Abbott.2017, Fragkoudi.2017, Parul.2020} and our analysis of the orbital structures in a few simulations with BP/X bulges (Tahmasebzadeh et al. 2024, in preparation); \textbf{(2) Classical bulge}: we assume that the classical bulge composed of nonperiodic box orbits with $\lambda_{z} < 0.4 $, which contribute to a hot, dispersion-dominated, and round structure in both face-on and edge-on views. It is widely accepted that classical bulges are slowly rotating and dominated by high-velocity dispersion \citep{Babusiaux.2010, Rojas-Arriagada.2014, Erwin.2021, Fragkoudi.2020} \textbf{(3) Nuclear disk}: a rotation-dominated structure perpendicular to the bar constructed by highly circular orbits with $\lambda_{z} > 0.4 $ in the inner regions; 
\textbf{(4) Main disk}:  all orbits with apocenter radii larger than the bar radius ($R_{\rm apo}>R_{\rm bar}$). It is dominated by highly circular orbits. Note that the orbits in the main disk extend well beyond the range of the available kinematic data, so they are constrained solely by the luminosity distribution of the galaxy.

We show the location of the four structures in the phase space of $\lambda_{z}$ vs. $r$ in the right panel of Fig. \ref{fig:cirty}.  Highly circular orbits distributed in the inner and outer regions make up the nuclear and main disks. The non-circular orbits could construct a classical bulge, and could also be part of the bar. The bar orbits overlap with the orbits of the classical bulge in the phase-space of $\lambda_z$ versus $r$. Overall, the structures are distinctly separated in the phase space distribution; however, the bar orbits in the inner regions exhibit partial overlap with orbits constituting the classical bulge in $\lambda_z$ versus $r$, illustrating the importance of detailed orbital classification using frequency analysis for separating the bar.


\begin{figure*}
	\centering	%
	\includegraphics[width=2\columnwidth]{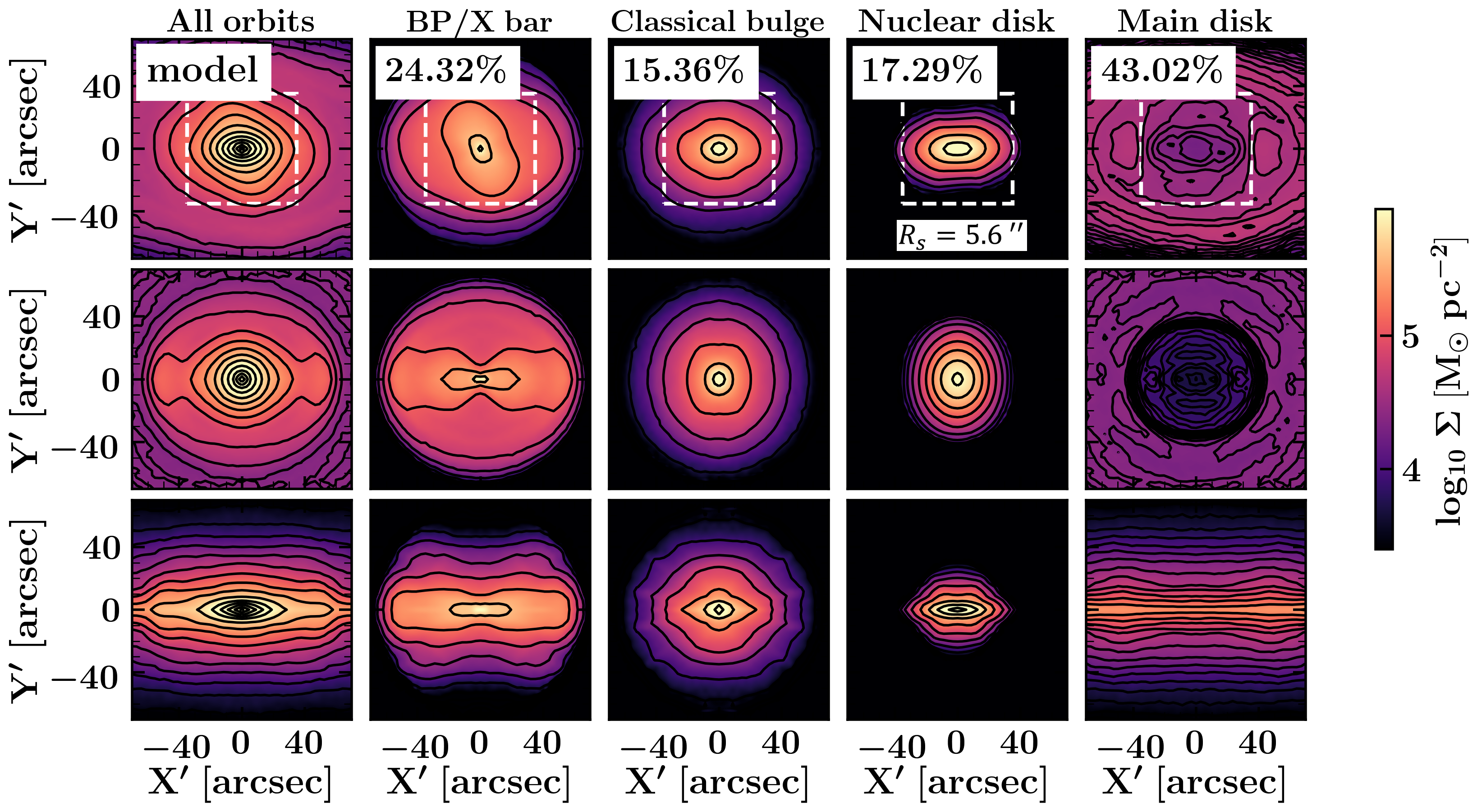}
	\caption{The surface density of different orbital structures decomposed from our best-fitting model constrained by TIMER data. The rows from top to bottom are projections with the observed galaxy orientation (top), face-on (middle), and edge-on (bottom) views. Columns from left to right show the reconstructed surface densities of the whole galaxy, the BP/X-shaped bar, the classical bulge, the nuclear disk, and the extended main disk, respectively. The white dashed squares in the first row indicate the data coverage. The luminosity fraction of each structure within the data coverage is tagged in the top row (in percent). The scale radius of the nuclear disk component is $R_{s}=5.6$ arcsec. }%
	\label{fig:decomposition}%
\end{figure*}

\begin{figure*}
	\centering	%
	\includegraphics[width=2\columnwidth]{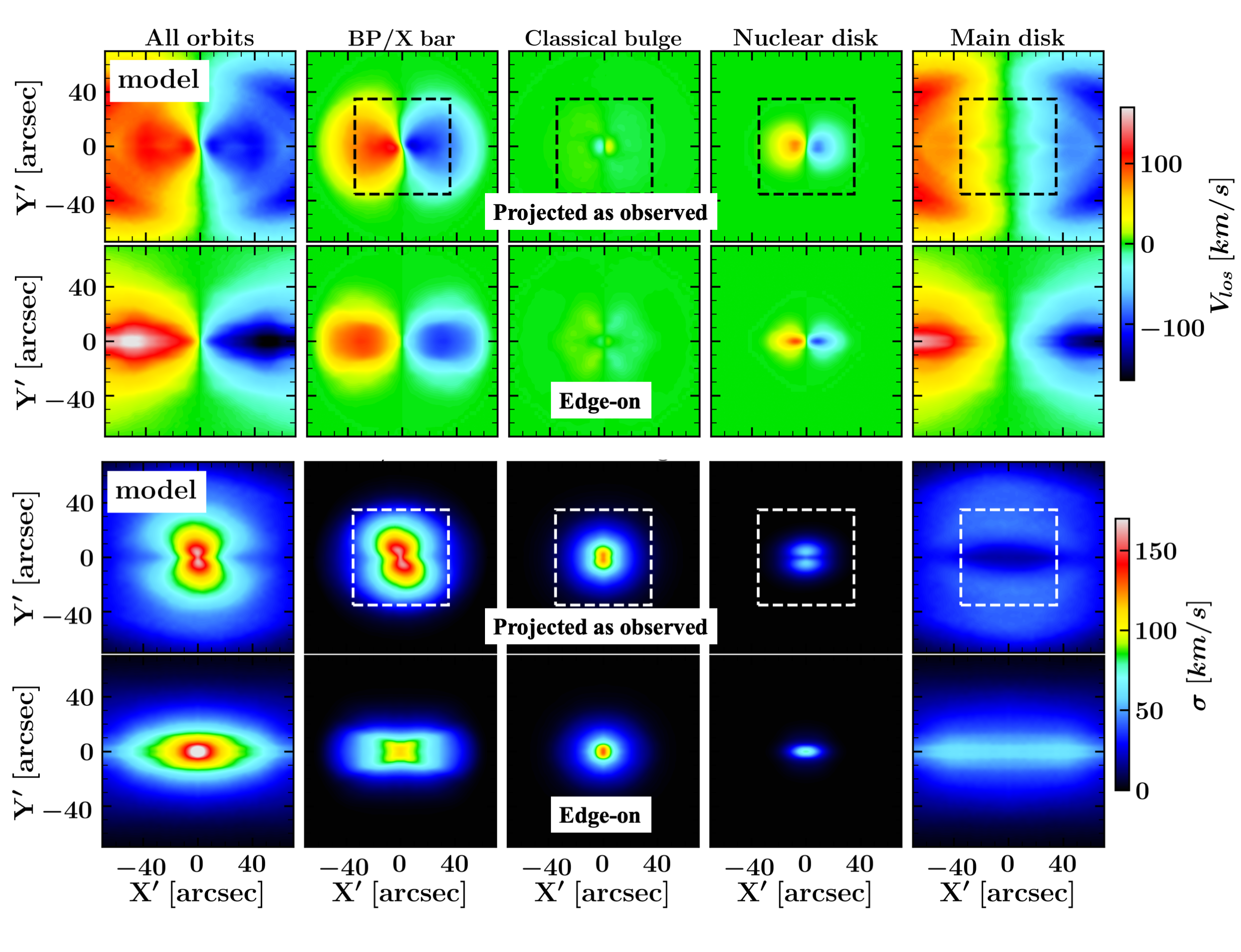}
	\caption{The kinematic maps of different structures decomposed from our best-fitting model. The first and the second rows show the velocity maps of the galaxy projected as observed and edge-on views. The third and fourth rows show the velocity dispersion maps of the galaxy projected as observed and edge-on views. Columns from left to right show the reconstructed kinematic maps of the whole galaxy, the BP/X-shaped bar, the classical bulge, the nuclear disk, and the extended main disk, respectively. The white dashed squares in the first column indicate the data coverage.}%
	\label{fig:decompositionV}%
\end{figure*}

\subsubsection{Morphology and kinematics of the four structures}
\subparagraph{Morphology of each structure:}

We reconstruct the 3D density distribution and kinematics by summing the particles sampled from the orbits, the entire galaxy from all orbits in the model and each structure from the orbits within each group. In Fig. \ref{fig:decomposition}, we display the surface densities of different components projected with the same orientation as the observed galaxy, in face-on and edge-on views, respectively. The columns from left to right represent the surface densities of the entire galaxy, BP/X bar, classical bulge, nuclear disk, and main disk orbits, respectively. The white square indicates the spatial coverage of the TIMER data. The surface density of the entire galaxy built by all orbits matches the global surface density from observations well. The variation of the PA derived from the entire galaxy rebuilt by the orbits, as obtained from the ellipse fitting, is presented in Fig. \ref{fig:PA}, and it is consistent with the PA variation for the GALFIT model and the S4G image shown in Fig. \ref{fig:photo}.
\par
As shown in Fig.\ref{fig:decomposition}, the bar exhibits BP/X shaped structure when viewing edge-on, the classic bulge is mostly round and might still be mixed with some flat disky orbits, the nuclear disk is elongated perpendicular to the bar when seen face-on and mostly flat when seen edge-on, the main disk is flat. 
Our results confirm that NGC 4371 has a BP/X bar that co-exists with a classical bulge, contributing nearly similar fractions to the total surface density. The BP/X-shaped structure is not directly discernible in the 2D image of this galaxy, and note that our projected 3D density model constructed in Section 3.1 lacks a BP/X-shaped structure where the bar is nearly prolate. Although a BP/X-shaped structure is not explicitly included in the gravitational potential, our model can still support orbits constructing the BP/X-shaped structure, and the kinematic constraints pick up the orbits contributed to the BP/X structure (see Fig.~\ref{fig:kinematic_PS}, Fig.~\ref{fig:decompositionV}). This capability is a key aspect of our model, as we discussed in detail in our previous works \cite{Behzad.2021, behzad.2022}. 

We then quantify the luminosity fraction of different structural components in the best-fitting model. The contributions of the four components to the total luminosity of the galaxy are $24.32\%$ BP/X bar, $15.36\%$ classical bulge, $17.29\%$ nuclear disk, and $43.02\%$ main disk. When considering only the region within the data coverage, these fractions change to $34.92\%$ BP/X bar, $25.69\%$ classical bulge, $30.23\%$ nuclear disk, and $9.16\%$ main disk, respectively. Note that we have flexibility in separating the classical bulge and inner disk. If we choose a difference of $0.1$ in the $\lambda_{z}$ cut-off, the classical bulge and nuclear disk fractions will change by $2\%$ to $5\%$.

In our model, the radially end-to-end separation of the X-shaped structure is approximately half of the bar length. This proportion is consistent with what is measured for the Galactic bar \cite[]{li.2012, Portail.2017a}. 
A barlens structure was reported in NGC 4371 \cite[]{Buta.2015}; however, we show that it is actually a nuclear disk in the central regions in agreement with \cite{Gadotti.2015}, and it does not contribute to the BP/X structure. \cite{Erwin.2015} employed 1D photometric decomposition to estimate the stellar mass ratio of the classical bulge to the total galaxy, finding it to be less than $10 \%$ for NGC 4371, which is lower than the fraction of the classical bulge we found dynamically.

\begin{figure*}
	\centering	%
	\includegraphics[width=1.7\columnwidth]{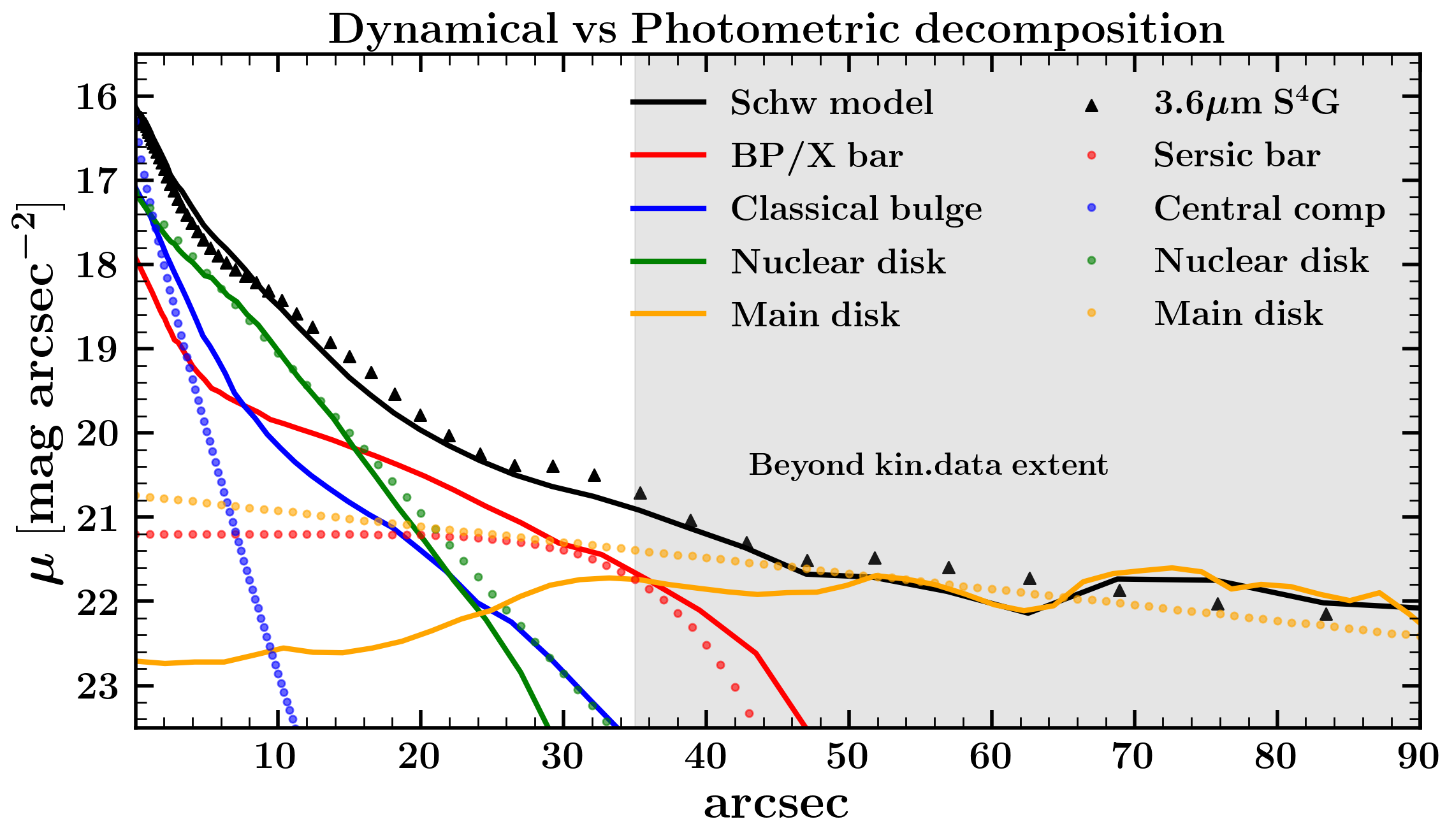}
	\caption{Comparison of dynamical structure decomposition with photometric decomposition. Solid lines are the 1D surface brightness profiles along the major axis for different orbital structures decomposed from our best-fitting model constrained by TIMER data. The dotted lines indicate those of the structures from the GALFIT decomposition of S$^4$G image. The shaded gray area represents regions beyond the kinematic data coverage.}%
	\label{fig:1Ddecomposition}%
\end{figure*}

\subparagraph{Kinematics of each structure:} we investigate the kinematic properties associated with these structures to validate the accuracy of our orbital decomposition. We show the LOS velocity (the first and second rows) and velocity dispersion (the third and fourth rows) of each component in Fig.\ref{fig:decompositionV}, by projecting them along two viewing angles: as the galaxy was observed and edge-on. 
In the maps projected as observed, the inner region of the BP/X bar shows the strongest regular rotation, which is caused by stellar motion through an X-like (or a $\infty$-like) path along the bar and moving inward/outward from the disk plane. The total velocity dispersion map displays two high-dispersion lobes in the vertical direction. This enhancement is caused by a combination of the BP/X structure and the classical bulge. In the BX/P bar, velocity dispersion increases in two wings of the X-shaped structure. 
In the map projected edge-on, the BP/X bar has moderate rotation, creating a relatively high dispersion in an X-shaped area. The classical bulge, characterized by its dynamically hot nature, contributes to an enhanced velocity dispersion at the centre. A weak counter-rotating motion in the central region of the classical bulge is due to orbits with $\lambda_z <0$. In contrast, the nuclear and main disks demonstrate strong regular rotation coupled with low dispersion.

\subparagraph{The 1D surface brightness profiles:} we extract the surface brightness profile along the major axis of the galaxy on the projected 2D observational plane for each component and compare the results of our dynamical decomposition and that of the photometric decomposition based on the S$^4$G image in Fig. \ref{fig:1Ddecomposition} (note that they are not necessary to be comparable, we do not impose any of these photometric decomposed components in our dynamical model rather than the global parametrised surface brightness of the galaxy). The surface brightness profiles of our four dynamically decomposed components, BP/X bar, classic bulge, nuclear disk, and main disk, can be roughly compared with the four photometrically decomposed components, the S\'{e}rsic bar, central compact component, nuclear disk, and main disk. The major difference is that the dynamical main disk exhibits a luminosity in the inner regions much lower than the inward extrapolation of an exponential profile from the outer regions. Thus, in the inner regions, the dynamical main disk contributes less luminosity than the photometric exponential main disk; instead, the dynamical BP/X bar and the classic bulge contribute more luminosity than the corresponding S\'{e}rsic bar and central compact component from photometric decomposition.
The main disks defined more physically from stellar dynamics or stellar populations are not necessarily exponential but have lower luminosity in the inner regions; this has been noted in both simulations and observations \cite[]{ling.1018, Breda.2020, Ding.2023}, and is consistent with our results here.
At the same time, the dynamical nuclear disk closely approximates an exponential distribution consistent with the photometric decomposition results. The scale radius of the nuclear disk, determined by dynamical decomposition, is $R_{s}=5.6$  arcsec, which is similar to $5.2$ arcsec obtained from photometric decomposition.

\section{Conclusions}\label{S:con}
We apply the triaxial Schwarzschild barred model presented in \cite{behzad.2022} to an S0 barred galaxy NGC 4371. The gravitational potential is adopted as the combination of a 3D stellar luminosity density multiplied by a constant stellar mass-to-light ratio, a spherical dark matter halo, and a fixed black hole. We use the 3D stellar luminosity density deprojected from a 2D photometric image, combining an axisymmetric disk, and a triaxial bar. We have five free parameters in the model: the stellar mass-to-light ratio $M_{*}/L$, dark matter halo mass $\log M_{200}/M_{*}$, the inclination angle of the disk $\theta$, the bar angle with respect to the intrinsic major axis of the disk $\varphi$, and the pattern speed $\Omega_{\rm p}$.

We independently created two sets of models constrained by stellar kinematic data from the TIMER and ATLAS3D surveys. For both sets, our models match the observational data well, capturing the major properties of the bar as seen in the observed 2D images and the stellar kinematic maps. The main results are as follows:
\par
(1) For the model using TIMER, we obtained $\theta = 60 \pm 2 ^\circ$, $\varphi = -12 \pm 1 ^\circ$, $M_{*}/L_{3.6 um}=1.05\pm 0.05$, and DM virial mass $\log_{10}(M_{200}/M_{*})=2.4 \pm 0.5$. For the model based on ATLAS3D data the best fitting parameters are $\theta = 60 \pm 1 ^\circ$, $\varphi = -12 \pm 1 ^\circ$, $M_{*}/L_{3.6 um}=0.99\pm 0.11$, and DM virial mass $\log_{10}(M_{200}/M_{*})=3.0 \pm 0.6$. The best-fitting parameters obtained from the two sets of models are generally consistent with each other, with larger uncertainty for ATLAS3D data with smaller data coverage than TIMER.
\par
(2) We constrain the pattern speed of NGC 4371 to be $\Omega_{\rm p} =  23.6 \pm 2.8 \hspace{.08cm} \mathrm{km \hspace{.04cm} s^{-1} \hspace{.04cm} kpc^{-1} }$ using TIMER data, and $\Omega_{\rm p} =  22.4 \pm 3.5 \hspace{.08cm} \mathrm{km \hspace{.04cm} s^{-1} \hspace{.04cm} kpc^{-1} }$ using ATLAS3D data. The bar pattern speed can be effectively constrained by the Schwarzschild model using data that cover only the bar region, consistent with previous results from validation exercises with mock data \citep{behzad.2022, Vasiliev.2019c, Dattathri.2023}. While it is necessary to have an observed velocity profile that covers both the bar and the disk outer region to accurately measure $\Omega_{\rm p}$ with the traditional TW method. The TW method relies solely on information about the LOS velocity. In contrast, the orbit superposition method utilizes four linearly independent moments of the LOSVD, which tightly constrain the allowed orbits in the modelled region, reflecting the overall kinematics of the bar and thereby enabling more accurate estimates of the bar pattern speed.
\par
(3) We determined the location of the corotation resonance, along with the bar pattern speed for each model. Using all our models within the $1\sigma$ confidence level, constrained by the TIMER data,  we found that the mean bar co-rotation radius for NGC 4371 could be around $131$ arcsec, with a standard deviation of $24$ arcsec.

\par
(4) We compute the dimensionless bar rotation parameter for each model considering the upper/lower limit of bar length of $ R_{\rm bar} = 69^{+6}_{-5} $ arcsec. In the best-fit Schwarzschild model, the bar rotation parameter is $ \mathcal{R} = R_{\rm cor}/R_{\rm bar}=2.5 \pm 0.2$, However, we report the bar rotation parameter for the galaxy by calculating the mean and standard deviation of all possible bar rotation parameter values across all models within the one-sigma region that results in $ \mathcal{R} = R_{\rm cor}/R_{\rm bar}=1.88 \pm 0.37$. This indicates that NGC 4371 likely has a slow bar with a lower limit close to the borderline.
\par
(5) We found a large amount of DM mass within the bar region. The fraction of dark matter to the total enclosed mass within the bar region ($\rm M_{\rm DM}/ M_{\rm total}$) is $\sim 0.51 \pm 0.06$. Our results align with cosmological simulations that suggest fast bars are typically found in baryon-dominated disks. However, to confirm the correlation between the bar rotation parameter and the amount of dark matter in the disk region for real galaxies, it is necessary to apply our modelling to a large sample of barred galaxies.
\par
(6) Using orbit classification, we dynamically decompose the galaxy into four components: BP/X bar, classical bulge, nuclear disk, and main disk, which contribute 34.92 $\%$, 25.69 $\%$, 30.23 $\%$, and 9.16$\%$ of the luminosity, respectively, within the MUSE data coverage of NGC 4371. We confirm that NGC 4371 has a BP/X bar and a classical bulge, consistent with kinematical features in velocity dispersion and $h_{4}$ maps \citep{Gadotti.2015}. The previously reported barlens structure \citep{Buta.2015} may actually be a nuclear disk component built by rotation-dominated orbits, which does not contribute to the BP/X bulge or the classical bulge.

We illustrate that our barred Schwarzschild model can reproduce the kinematic properties of real barred galaxies. It is a powerful new tool for uncovering key properties of the barred galaxies, including the bar pattern speed and the internal BP/X-shaped orbital structure. 
This framework will be included as a module in the publicly available DYNAMITE page \cite[]{Jethwa.2020,Sabine.2022}, which is a new implementation of the code by \cite{bosch.2008}. This methodology will be applied to large samples of barred galaxies from different mass ranges and environments. It will help us to understand the formation and evolution of barred galaxies by investigating their pattern speed, BP/X structures, black holes, and dark halo masses.  

\section*{Acknowledgements}
LZ acknowledges the support from the National Natural Science Foundation of China under grant No. Y945271001, and the CAS Project for Young Scientists in Basic Research under grant No. YSBR-062.  BT and MV
gratefully acknowledge funding from the National Science Foundation (grants NSF-AST-2009122). This work was supported by STFC grants ST/T000244/1 and ST/X001075/1. AdLC acknowledges support from Ministerio de Ciencia e Innovación through the Spanish State Agency (MCIN/AEI) and from the European Regional Development Fund (ERDF) under grant CoBEARD (reference PID2021-128131NB-I00), and under the Severo Ochoa Centres of Excellence Programme 2020-2023 (CEX2019-000920-S). Based on observations collected at the European Southern Observatory under ESO program 060.A-9313(A). J.M.A. acknowledges the support of the Viera y Clavijo Senior program funded by ACIISI and ULL and the support of the Agencia Estatal de Investigaci\'on del Ministerio de Ciencia e Innovaci\'on (MCIN/AEI/10.13039/501100011033) under grant nos. PID2021-128131NB-I00 and CNS2022-135482 and the European Regional Development Fund (ERDF) ‘A way of making Europe’ and the ‘NextGenerationEU/PRTR.

\section*{DATA AVAILABILITY}
The raw and reduced MUSE data used in this work are publicly available at the ESO Science Archive Facility.

	
	
\bibliographystyle{mnras}
\bibliography{library}


\newpage	


\appendix
\setcounter{figure}{0} 
\renewcommand{\thefigure}{A\arabic{figure}} 

\section{Axisymmetric Model} \label{Appendix}
We also considered a nearly axisymmetric model by fitting MGEs without twists ($ \Delta \psi_{j}^{\prime} = 0 $) to the whole galaxy. Details of fitting are shown in Table \ref{table:allmgep}. For the nearly axisymmetric model, we explore intrinsic shape parameters $ \sigma_{j}, p_{j}$, and $q_{j}$ that are more efficient than searching over the viewing angles. $p_{j}$ and $q_{j}$ are the intermediate-to-long and short-to-long axis rations of the Gaussian component $j$. The deprojection of an axisymmetric system cannot constrain the $\varphi$ as it is irrelevant but has a finite axis-ratio between $y$ and $x$. We allow for some degree of triaxiality of the galaxy by setting a non-unity $u$. The stationary axisymmetric model also has five free hyperparameters: $M_{*}/L$, $q$, $p$, $u$, and $M_{200}/M_{*}$. We use an interval of $ 0.01$ for $q$, $p$, and $u$ in the searching process.
\par
We sample a set of tube orbits in $ x-z $ plane and box orbits from equipotential surfaces with zero velocity in the energy $E$, which are discretized by spherical angles of $\Theta$ and $\Phi$. The number of start points of the box orbit library $(E, \Theta, \Phi)$ is also $30 \times 15 \times 13$. To reduce the Poisson noise of the model, we consider the dithering number to be 3, so each orbital bundle contains $27$ orbits with close starting points.  
\par
The best-fitting axisymmetric model and its comparison versus the best-fitting bar model is shown in Fig. \ref{fig:axi_bar}. Fig. \ref{fig:Axi_analysis} shows the main differences in recovered properties between the axisymmetric model versus the bar model, including variation of the axial ratios (left panel), enclosed mass profiles (middle panel), and the stellar orbit distribution in the space of circularity $\lambda_{z}$ vs. time-averaged radius $r$ (right panel). Although the axisymmetric model apparently matches the observation well, there are significant differences in the internal orbital structure and mass profile of the best-fitting axisymmetric model compared to the bar model. The improvement of the bar model is evident in the velocity residual map (third row, second column) when compared to the axisymmetric model (last row, second column). This improvement is attributed to the model's ability to capture the higher stellar velocities along the bar. While producing a higher stellar motion, the axisymmetric model prefers the best-fitting model with a larger inclination ($\theta = 75^{\circ}$), which leads to a significantly thicker disk as is shown in the left panel Fig. \ref{fig:Axi_analysis}, $q$ in the disk region is $\sim 0.6$ for axisymmetric model (green solid line) and $\sim 0.4$ for the bar model (green dashed line). Previous studies measured the inclination angle for this galaxy to be $\sim 60$ \cite[]{Erwin.2008,Gadotti.2015} similar to our bar model.

\begin{figure}
	\centering	%
	\includegraphics[width=0.7\columnwidth]{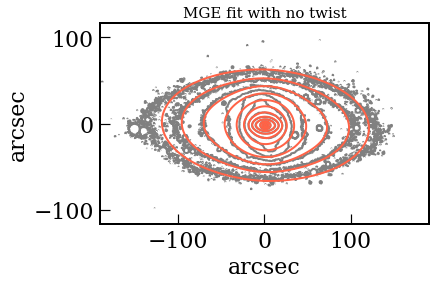}
	\caption{The contours of the NGC 4371 S$^4$G image (grey), overplotted with contours of the best-fitting MGE (red) without twist between different Gaussians applied to the whole galaxy.}%
	\label{fig:all_MGE}%
\end{figure}

\begin{table}
	\centering
	\footnotesize
	\begin{tabular}{p{0.15\linewidth}p{0.15\linewidth}p{0.15\linewidth}p{0.15\linewidth}p{0.15\linewidth}}
		\hline
		$ j $    &  $ L_{j}$ $(\mathrm{L_{\odot}pc^{-2}}) $ &  $ \sigma^{\prime}_{j} $ $ (\mathrm{arcsec}) $    & $ q^{\prime}_{j} $   &  $ \Delta \psi_{j}^{\prime} $ $(^{\circ})$   \\
		\hline
        $1$  & 50033.382 & 1.35 & 0.74 & 0.0  \\
        $2$ & 5211.781  & 7.874 & 0.53 & 0.0  \\
        $3$ & 1104.101  & 11.242 & 0.95 & 0.0  \\
        $4$ & 616.274  & 19.265 & 0.95 & 0.0  \\
        $5$ & 209.024  & 54.744 & 0.53 & 0.0  \\
        $6$ & 38.302   & 109.426 & 0.53 & 0.0  \\			
		\hline
	\end{tabular}
	\\
	\parbox{\columnwidth}{\caption{Details of MGE fit for the whole galaxy in axisymmetric limit. $ j $ is the number of each individual Gaussian for which,  $ L_{j}$ is the central flux in the unit of $(\mathrm{L_{\odot}pc^{-2}})$, $ \sigma^{\prime}_{j} $ presents the size in the unit of $ (\mathrm{arcsec}) $, $q^{\prime}_{j}$ indicates the flattening. $ \Delta \psi_{j}^{\prime} $ is twisted angle of Gaussian that is $0$ in this case. }
		\label{table:allmgep}}
\end{table}


 \begin{figure*}
	\centering	%
	\includegraphics[width=2\columnwidth]{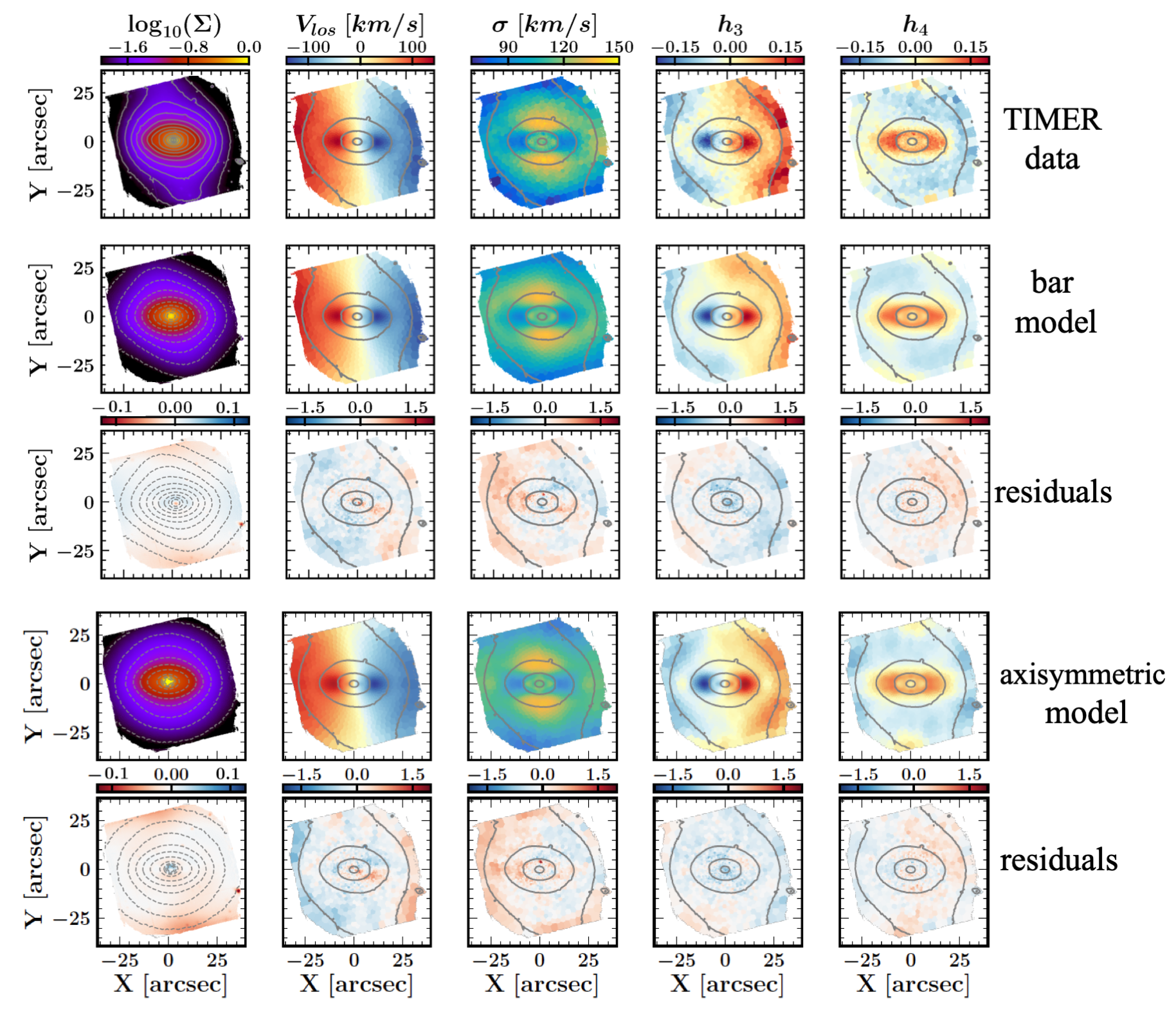}
	\caption{Comparison of the best-fitting bar model obtained with $\theta = 59^{\circ}$ (second row) versus the best-fitting axisymmetric model (fourth row) obtained with $\theta = 75^{\circ}$ using the TIMER data (first row). The third and fifth rows show the residual of each model. The surface density and its residuals are presented in the first column, for TIMER data, we use the flux of the IFU data cube and its contours, while for the bar and axisymmetric models, we used the MGEs surface density and its counters of the best-fitting models.}%
	\label{fig:axi_bar}%

\end{figure*}

 \begin{figure*}
	\centering	%
	\includegraphics[width=2\columnwidth]{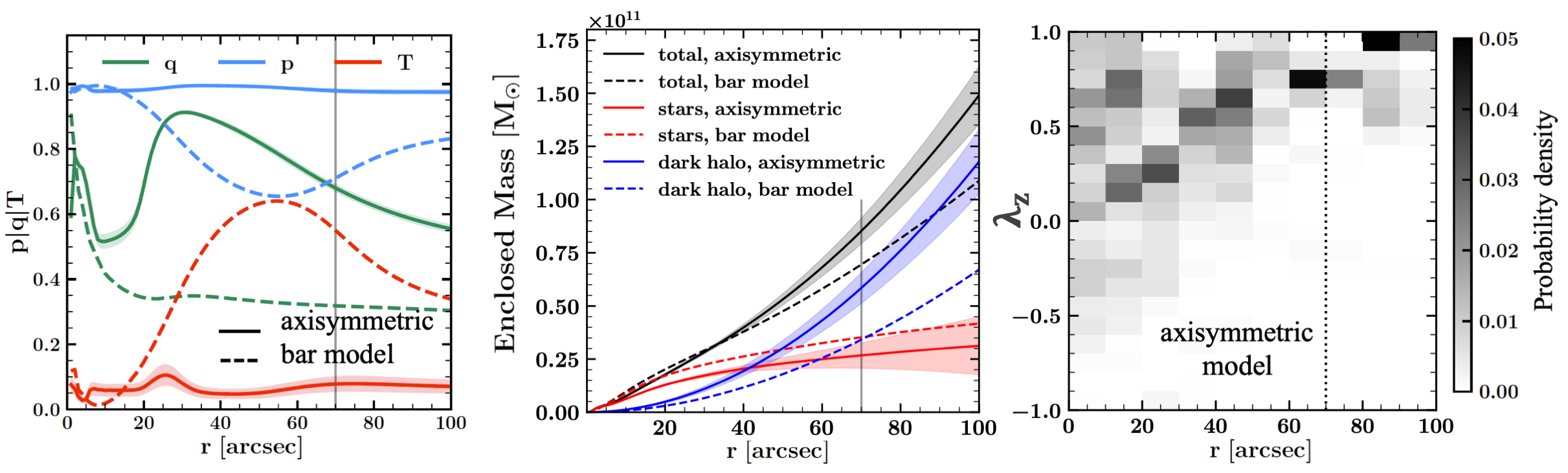}
	\caption{Left panel:  variation of the axial ratios $q = c/a$ (green), $p = b/a$ (blue), and triaxial parameter $T = (1 - p^{2})/(1 - q^{2})$ (red) for the best-fitting axisymmetric model (solid lines) compared with the bar model (dashed lines) using TIMER.  Middle panel:  mass profiles of the best-fitting axisymmetric model (solid lines) compared with the bar model (dashed lines). The red, blue, and black curves represent the stellar mass, dark matter mass, and total mass, respectively. The shaded regions indicate the $1 \sigma$ uncertainty from our axisymmetric models. The right panel represents the stellar orbit distribution in the space of circularity $\lambda_{z}$ vs. time-averaged radius $r$ for the best-fitting axisymmetric model with TIMER. The vertical dashed lines indicate the kinematics data extent.}%
	\label{fig:Axi_analysis}%
\end{figure*}

 \begin{figure*}
	\centering	%
	\includegraphics[width=2.0\columnwidth]{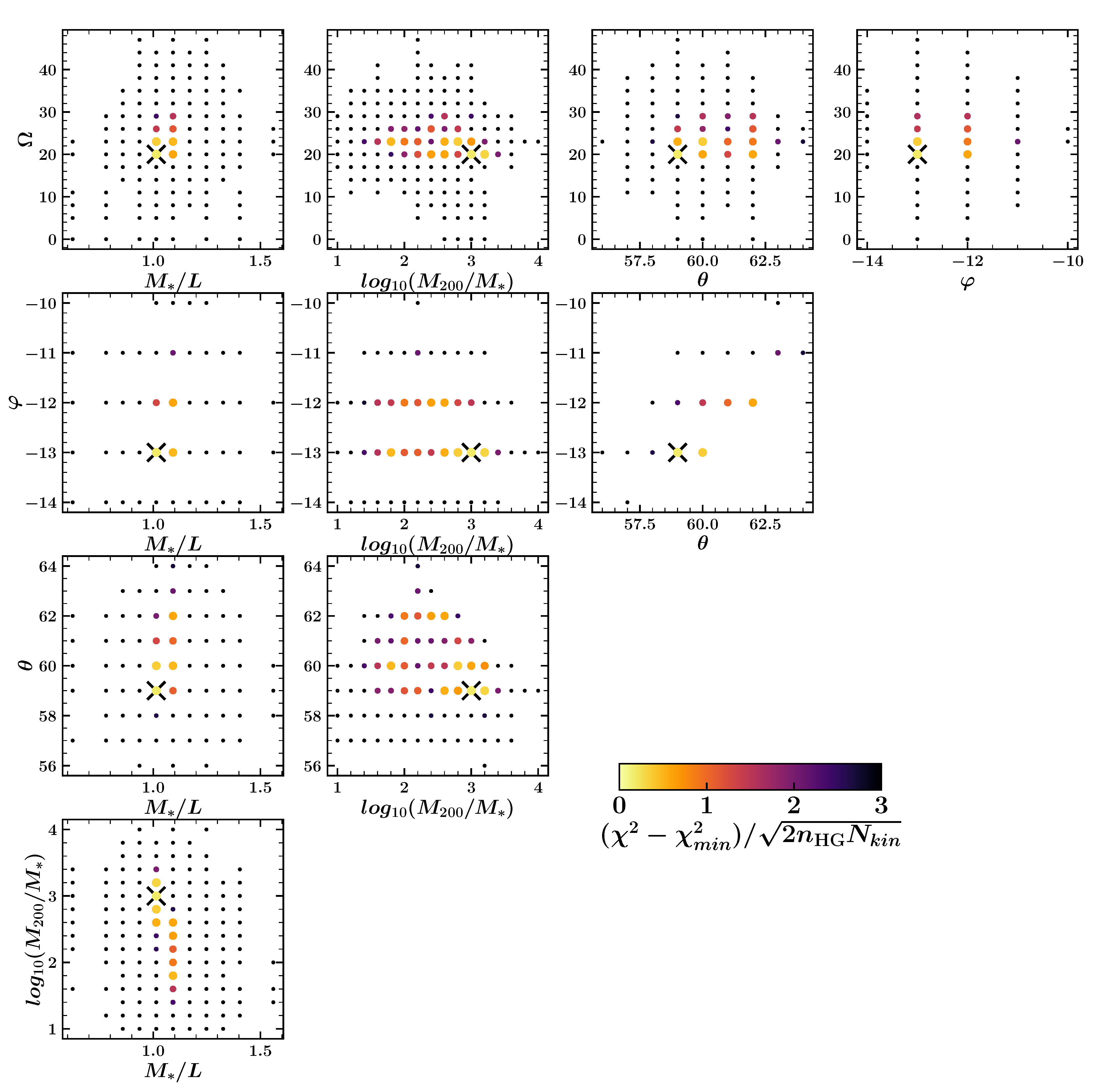}
	\caption{The parameter grid we explored for the bar model
          fitting using TIMER data. The five hyperparameters are stellar mass-to-light
          ratio $M_{*}/L_{3.6 um}$ in unit of $M_{\odot}/L_{\odot}$, dark matter halo mass $\log
          M_{200}/M_{*}$, the inclination angle of the disk $\theta$, the bar angle with respect to the major axis of the disk $\varphi$, and the pattern speed $\Omega_{\rm p}$ in units of $\mathrm{km \hspace{.04cm} s^{-1} \hspace{.04cm} kpc^{-1} }$. Each point is one model color-coded according to their $\chi^{2}$ values shown in the color bar. 
          The points with $(\chi^{2}-\chi^{2}_{min})/\sqrt{2n_{\rm GH}N_{\rm kin}}<1$
          indicate models within $1 \sigma$ confidence level. The black crosses
          indicate the best-fitting model.}%
	\label{fig:muse_bar_chi2}%
\end{figure*}

 \begin{figure*}
	\centering	%
	\includegraphics[width=2.0\columnwidth]{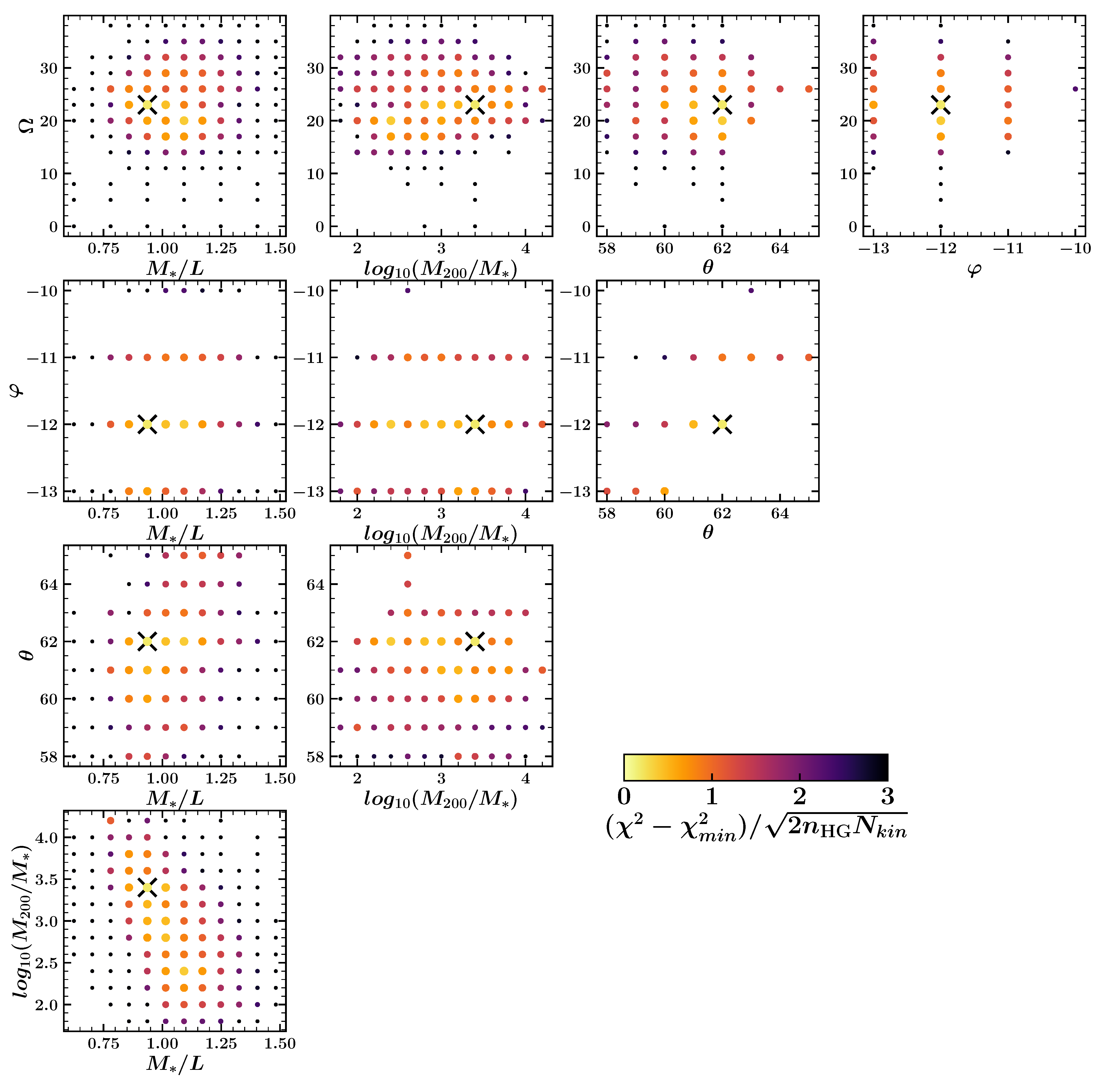}
	\caption{Same as Fig. \ref{fig:muse_bar_chi2} but for the bar model
          fitting using ATLAS3D data}
	\label{fig:atlas_bar_chi2}%
\end{figure*}

 \begin{figure*}
	\centering	%
	\includegraphics[width=2.0\columnwidth]{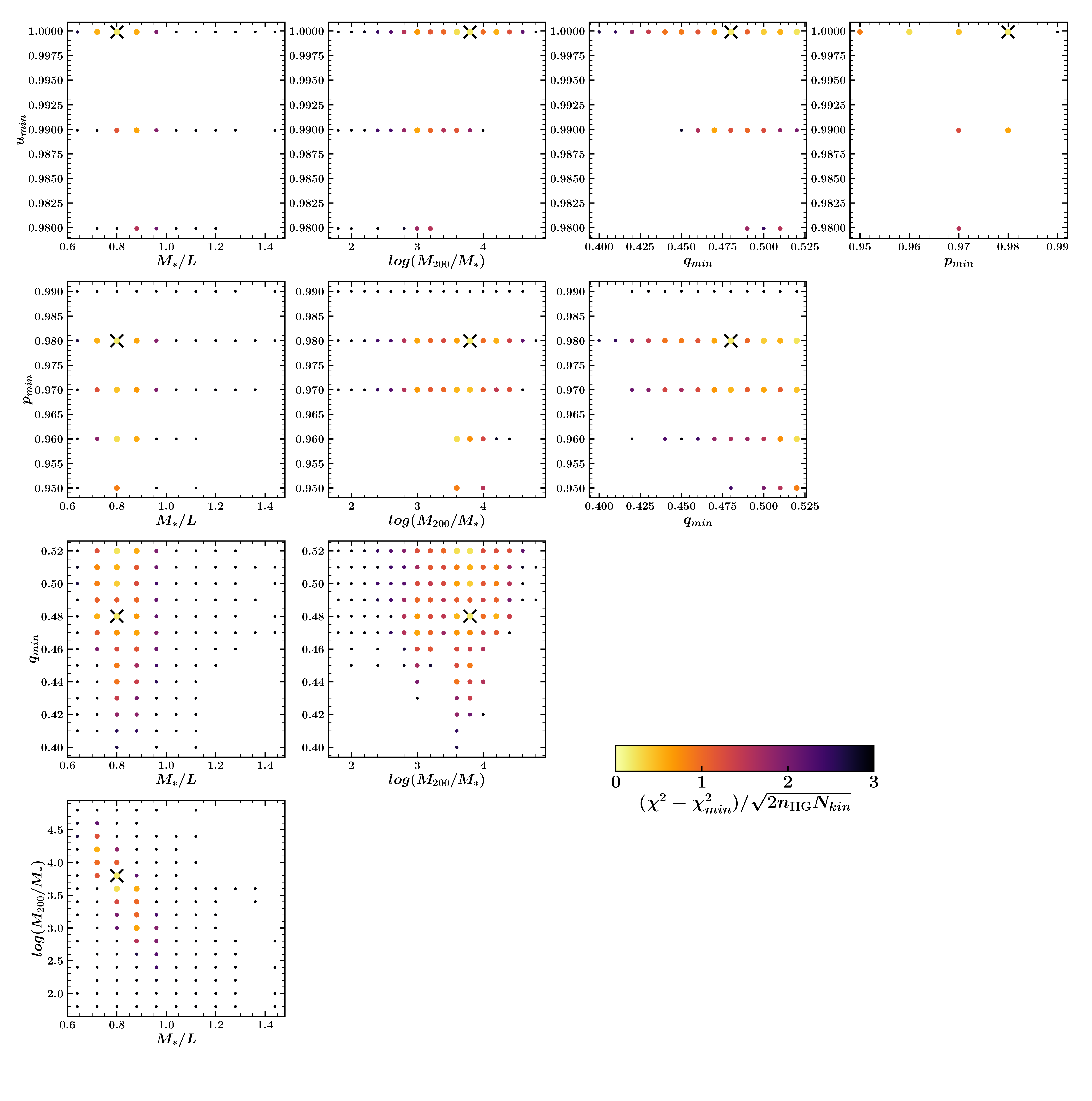}
	\caption{Same as Fig. \ref{fig:muse_bar_chi2} but for the axisymmetric model
          fitting using TIMER data. For the axisymmetric modeling, we use intrinsic parameters of $q,p,$ and $u$ instead of viewing angles of $\theta, \varphi,$ and $\psi$.} %
	\label{fig:muse_axi_chi2}%
\end{figure*}

 \begin{figure}
	\centering		\includegraphics[width=1.0\columnwidth]{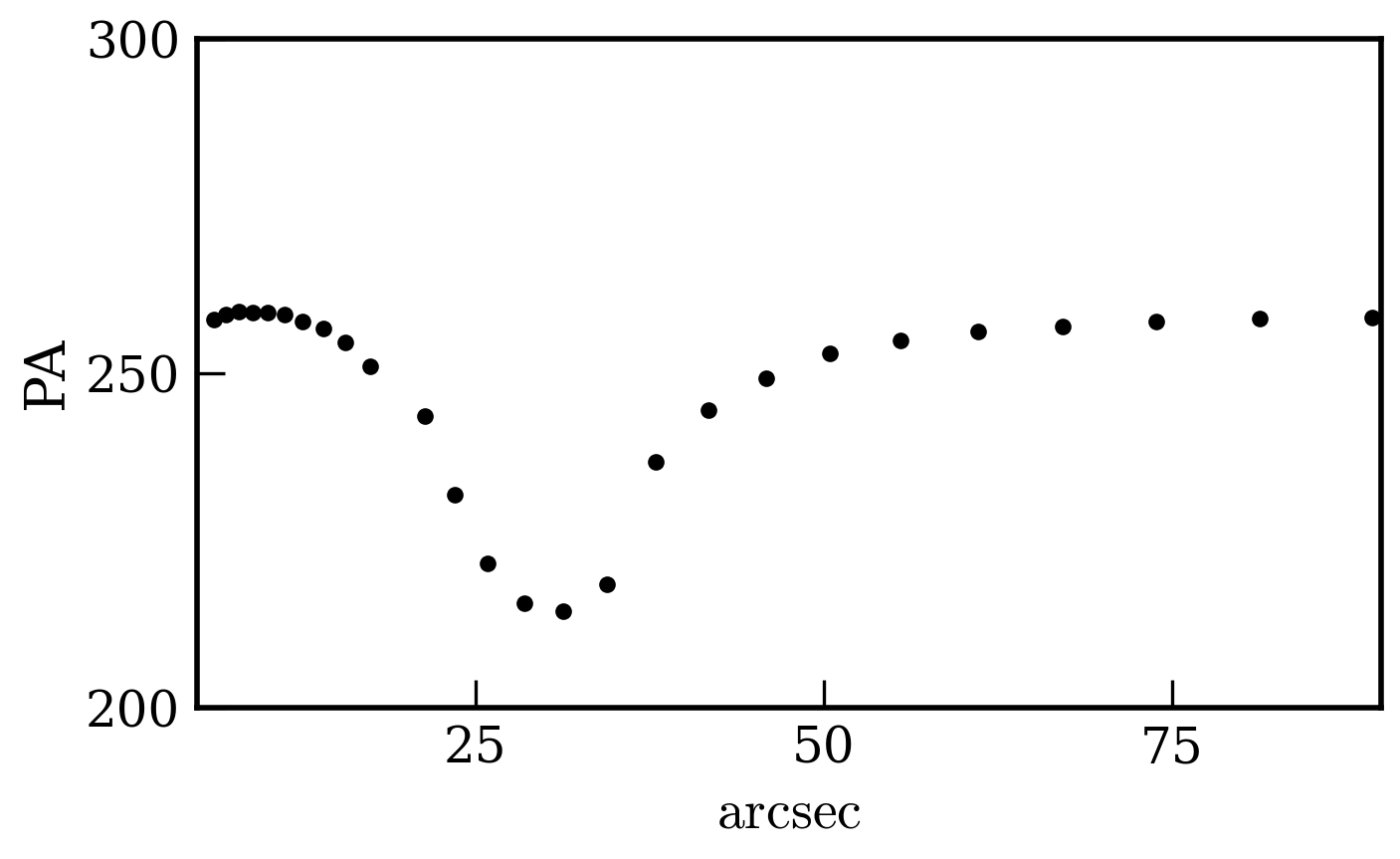}
	\caption{The variation of PA in the best fitting Schwarzschild model was extracted using ellipse fitting, similar to the method applied in Fig. \ref{fig:photo} for the GALFIT model and the S4G image, and overall is consistent with them.} %
	\label{fig:PA}%
\end{figure}

\clearpage

\bsp	
\label{lastpage}

\end{document}